\documentclass{aa} 
\usepackage{graphicx}
\usepackage{txfonts}
\begin{document} 

\title{The population of hot subdwarf stars studied with Gaia}
\subtitle{IV. Catalogues of hot subluminous stars based on Gaia EDR3 \thanks{The catalogues are only available in electronic form at the CDS via anonymous ftp to cdsarc.u-strasbg.fr (130.79.128.5) or via http://cdsweb.u-strasbg.fr/cgi-bin/qcat?J/A+A/}}

\author{R.~Culpan \inst{1}
    \and S.~Geier \inst{1}
    \and N.~Reindl \inst{1}
    \and I.~Pelisoli \inst{2}
    \and N.~Gentile Fusillo \inst{3}
    \and A.~Vorontseva \inst{4}}

\offprints{R.\,Culpan,\\ \email{rick@culpan.de}}

\institute{Institut f\"ur Physik und Astronomie, Universit\"at Potsdam, Haus 28, Karl-Liebknecht-Str. 24/25, D-14476 Potsdam-Golm, Germany
\and Department of Physics, University of Warwick, Coventry, CV4 7AL, UK
\and European Southern Observatory, Karl Schwarzschild Straße 2, 85748 Garching, Germany
\and Sixfold GmbH, Magirus-Deutz-Straße 16, 89077 Ulm, Germany}

\date{Received 16/02/2022 \ Accepted 07/03/2022}

\abstract
{In light of substantial new discoveries of hot subdwarfs by ongoing spectroscopic surveys and the availability of the {\em Gaia} mission Early Data Release 3 (EDR3), we compiled new releases of two catalogues of hot subluminous stars: The data release 3 (DR3) catalogue of the known hot subdwarf stars contains 6,616 unique sources and provides multi-band photometry, and astrometry from {\em Gaia} EDR3 as well as classifications based on spectroscopy and colours. This is an increase of 742 objects over the DR2 catalogue. This new catalogue provides atmospheric parameters for 3,087 stars and radial velocities for 2,791 stars from the literature. In addition, we have updated the {\em Gaia} Data Release 2 (DR2) catalogue of hot subluminous stars using the improved accuracy of the {\em Gaia} EDR3 data set together with updated quality and selection criteria to produce the {\em Gaia} EDR3 catalogue of 61,585 hot subluminous stars, representing an increase of 21,785 objects. The improvements in {\em Gaia} EDR3 astrometry and photometry
compared to {\em Gaia} DR2 have enabled us to define more sophisticated selection functions. In particular, we improved hot subluminous star detection in the crowded regions of the Galactic plane as well as in the direction of the Magellanic Clouds by including sources with close apparent neighbours but with flux levels that dominate the neighbourhood.} 


\maketitle

\section{Introduction \label{sec:intro}}

The name `subdwarf' was suggested in 1939 by \citet{kuiper39} to describe the stars found between the main sequence and the white dwarf regions of the Hertzsprung-Russell diagram (HRD). Most of those cool objects of late spectral types were later identified as metal-poor main sequence stars from old Galactic populations. The first hot subdwarf stars with early spectral types were found somewhat later by \citet{humason47} and in contrast to their cool siblings, their formation mechanism and evolutionary status are still unclear. 

\begin{figure}
  \centering
  \includegraphics[width=\hsize]{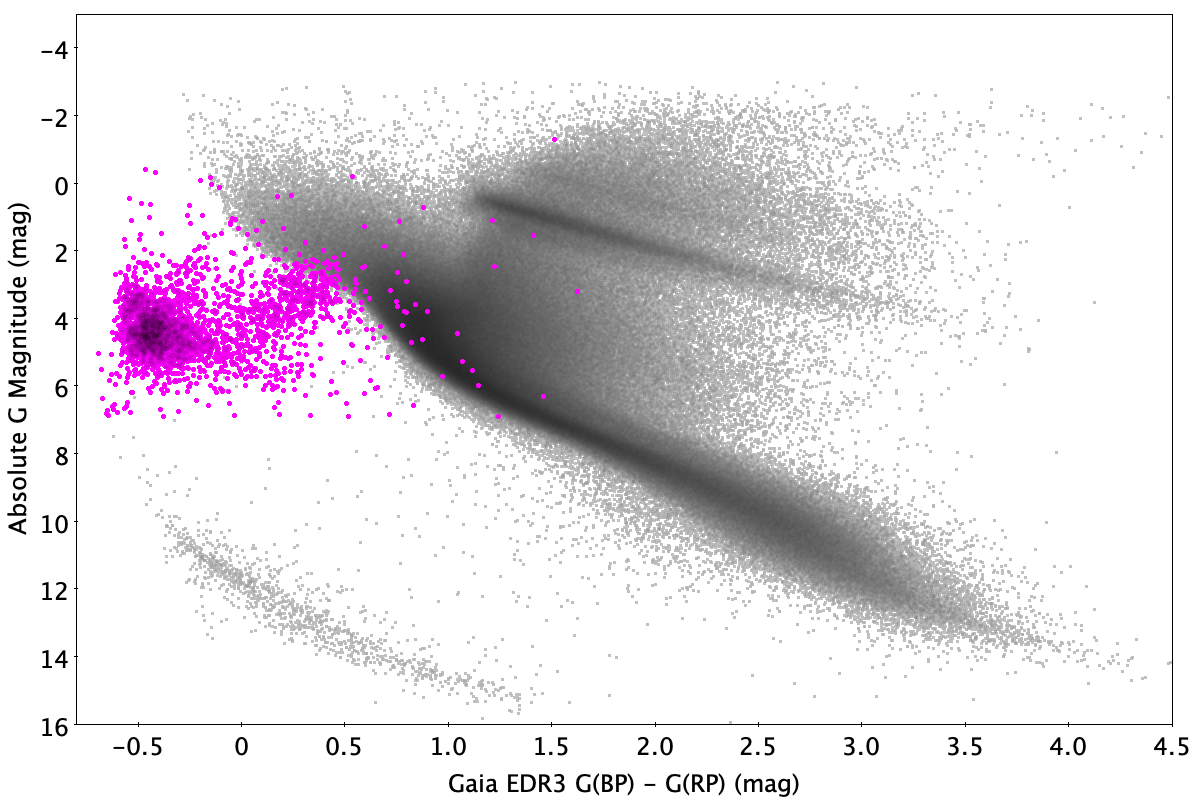}
 \caption{Full-scale {\em Gaia} EDR3 colour--magnitude diagram (CMD) (grey dots) with the known hot subdwarfs from Geier (2020) (magenta circles) highlighting the {\em Gaia} EDR3 CMD region where hot subdwarfs are expected. Here we can see the EHB cluster of sources at $(G_{\rm BP} - G_{\rm RP}) \sim -0.35$, $G_\textrm{abs} \sim 4.5$ and the extended cloud of hot subdwarf plus main sequence binaries towards $(G_{\rm BP} - G_{\rm RP}) \sim 0.5$, $G_\textrm{abs} \sim 3.0$.}
  \label{cmd_gaia_full_known_dr2}
  \end{figure}

Many, but not all hot subdwarfs lie at the extreme blue end of the horizontal branch (extreme horizontal branch (EHB)) in the HRD \citep{heber86d} (see Figure~\ref{cmd_gaia_full_known_dr2}). Only very few stars are predicted by single star evolution in this region as the evolutionary timescale of the post-asymptotic giant branch (post-AGB) phase is very short. Many hot subluminous stars are therefore likely to be helium-burning stars with longer evolutionary timescales. To evolve to the EHB, red giants or subgiants must lose almost their entire hydrogen envelope, which is explained by various scenarios of binary merger or mass transfer \citep{han02,han03}. See \citet{heber16} for a review. More recently, this region in the HRD was found to also contain cooling progenitors of helium white dwarfs \citep{heber2003} and other rare types of detached or accreting compact binary systems.

Initially, hot subdwarf stars were discovered via photometric surveys of faint blue stars \citep[e.g.][]{humason47,luyten51,luyten53,iriarte57,feige58,chavira58,chavira59,haro62,slettebak71,berger84,green86,demers86,downes86}. \citet{kilkenny88} published the first catalogue of 1225 spectroscopically identified hot subdwarf stars. See \citet{lynas04} for a review of those early studies. Surveys mainly targeting extragalactic sources detected many more such stars \citep{hagen95,wisotzki96,stobie97,mickaelian07,mickaelian08} and \citet{oestensen06} compiled a database containing more than 2300 entries. 

In the following decade, the Sloan Digital Sky Survey (SDSS) provided spectra of almost 2000 hot subdwarfs \citep{geier15b,kepler15,kepler16,kepler19} and new samples of bright hot subdwarf stars were observed \citep{vennes11}. In parallel, more and more data from new large-area photometric and astrometric surveys were gathered in multiple bands from the ultraviolet to the far-infrared. The availability of this amount of high-quality data triggered the first data release of the catalogue of known hot subdwarf stars \citep{geier17a}. 

In the following years, substantial new discoveries were made of hot subdwarfs and quantitative spectroscopic analyses of large samples have since been conducted in the course of ongoing spectroscopic surveys such as LAMOST \citep{lei18,lei19,lei20,luo19,luo21} and SDSS \citep{geier17b, kepler19}. A major milestone was the publication of {\em Gaia} mission DR2 \citep{gaia18}, because it allowed us to use colour, absolute magnitude, and reduced proper motion criteria to identify 268 previously misclassified hot subdwarfs as white dwarfs, blue horizontal branch (BHB), and main sequence stars. The release of the second version of the catalogue \citep{geier20} contained 5874 unique sources, provided atmospheric parameters for 2187 stars, and radial velocities for 2790 stars from the literature. This added 528 newly discovered hot subdwarfs to the \citet{geier17a} catalogue (see Figure~\ref{cmd_gaia_full_known_dr2}).

The \citet{geier17a} catalogue has been used to determine selection criteria for an all-sky catalogue of hot subdwarf star candidates selected from {\em Gaia} DR2 \citep{gaia18} by means of colour, absolute magnitude, and reduced proper motion cuts. This catalogue contains 39\,800 unique sources, has a magnitude limit of $G<19\,{\rm mag,}$ and covers the whole sky with the exception of the Galactic plane and the direction of the Large and Small Magellanic Clouds (LMC/SMC). It should be fairly complete up to $\sim1.5\,{\rm kpc}$ and the contamination by cooler stars should be about $10\%$ \citep{geier19}.

Here we present new releases of the catalogue of known hot subdwarfs (DR3) and the catalogue of hot subluminous star candidates (DR2) based on improved data from {\em Gaia} EDR3 \citep{gaia21}, new results from spectroscopic surveys, and an extensive literature search. The catalogue of known hot subdwarfs contains stars of spectral types O to B occupying the region in the colour--magnitude diagram between the more luminous main sequence stars of similar spectral type and the less luminous white dwarfs and their cooling curve. In addition, it also includes the known hot subdwarfs in binary systems with main sequence companions. The aggregated colour and absolute magnitude of these unresolved binaries is found between those of the individual stars causing the large spread seen in Figure~\ref{cmd_gaia_full_known_dr2}. 

The {\em Gaia} EDR3 catalogue of hot subluminous stars contains candidates in the same region of the HRD. Next to single subdwarfs and composite subdwarf binaries, it also contains subluminous stars of A and early-F type. 

The {\em Gaia} EDR3 hot subluminous star catalogue and the catalogue of known hot subdwarfs have been brought together to examine the possible effects of binarity on the {\em Gaia} EDR3 photometric and astrometric measurements and their associated errors. Differences have been found that correlate with the type of binary system that prevails.

The contents of this paper are as follows: Section 2  provides a catalogue of the known hot subdwarfs: Data Release 3. In Section 3 we describe how we constructed the hot subluminous star catalogue from {\em Gaia} EDR3. In Section 4 we consider astrometric and photometric indications of variability. Section 5 outlines the main specifications of the catalogues in terms of magnitude, distance and sky coverage. Section 6 contains a summary and the conclusions drawn.

%--------------------------------------------------------------------
\section{Catalogue of known hot subdwarfs: Data Release 3} 

\subsection{Input data}

We complement the spectroscopically classified hot subdwarfs from the \cite{geier20} catalogue with 224 newly discovered objects from DR6/7 of the LAMOST survey \citep{luo21}. These latter authors also provide atmospheric parameters and radial velocities for more than $1500$ hot subdwarfs observed in all the available releases of LAMOST. The second main source of new and refined classifications of 107 helium-rich hot subdwarf stars and related objects is provided by a targeted survey conducted with the South African Large Telescope (SALT) by \citet{jeffery21}. The discovery of a large sample of eclipsing HW\,Vir type binaries with hot subdwarf primaries has been reported by \citet{schaffenroth19}. Those that have been confirmed spectroscopically have been added to the catalogue. In contrast to the previous releases, we decided to also include the 196 spectroscopically identified hot subdwarfs in the globular clusters M80 \citep{monibidin09}, $\omega$\,Cen \citep{monibidin12,latour18b,monibidin09}, and NGC\,6752 \citep{heber86a,monibidin07}.

A further extension of our catalogue is provided by the rare spectroscopically identified stars in the post-AGB region and central stars of planetary nebulae (CSPNe), which are situated below the MS and have not yet evolved to the WD stage. Thus, they also belong to the population of the hot subdwarfs. The spectroscopically identified CSPNe are compiled in the catalogue of \citet{weidmann20}. From this sample, we selected CSPNe with classifications as O-type or emission line stars excluding both hot WDs and CSPNe located on or above the MS in the CMD. Those are regarded as the best candidates for hot subluminous stars, but as no clear distinction is possible, the catalogue of \citet{weidmann20} should be regarded as complimentary to our catalogue.

Adding the known post-AGB stars without nebulae, we find a total of 402 such objects in the literature \citep{rauch95,dreizler98,napiwotzki99,werner04a,werner04b,werner05,werner14,herald11,ziegler12,reindl14,reindl16,reindl17,reindl20,aller15,demarco15,werner15,hillwig17,loebling20}. The catalogue now includes, for the first time, all spectroscopically confirmed hot subdwarf stars regardless of their Galactic population and their evolutionary stage.

In addition, several recent studies reported the discovery of individual subdwarfs \citep{hillwig17,kupfer17a,kupfer17b,kupfer20,kupfer22,holdsworth17,ratzloff19,kilkenny19,bell19,hogg20,ratzloff20,silvotti21,pelisoli21,vos21,dorsch22,silvotti22,werner22a,werner22b}, all of which are included in our compilation.

\subsection{Astrometry and multi-band photometry}

{\em Gaia} EDR3 \citep{gaia21} astrometry and photometry has been added to the catalogue. The zero-points of the parallax measurements have been corrected as described in \citet{lindegren21} and the corrected {\em Gaia} parallaxes are provided in addition to the uncorrected ones. 

The sources for the further multi-band photometry as well as the Galactic extinction and reddening provided in the catalogue are described in \citet{geier20}. Only the infrared photometry in the $YJHK$ bands from the Visible and Infrared Survey Telescope for Astronomy (VISTA) Hemisphere Survey has been updated to the most recent DR5 \citep{mcmahon13}.

\subsection{Cleaning the catalogue}\label{cleaning}

To remove misclassified objects, we used colour indices as described in \citet{geier20}. As most of the newly discovered subdwarfs have been well classified and analysed, we did not find any cool outliers in this data release. Due to the PNs surrounding them and the fact that many of them are found at low latitudes and high reddening, we did not apply any colour cleaning to the CSPNe. Calculating the absolute magnitudes of all stars with accurate parallaxes (parallax error smaller than $20\%$ and zero-point correction applied), we identified several misclassified white dwarfs (WDs) with $M_{\rm G}<6.5\,{\rm mag}$ and also brighter objects such as main sequence B (MS-B) stars. The reduced proper motion $H=G+5\log{\mu }+5$ was calculated for stars with parallax errors $\geq20\%$ using the {\em Gaia} G magnitudes and proper motion ($\mu$) to identify additional WDs with $H>15$. However, we find no additional misclassified objects. 

Several misclassified subdwarfs were found by cross-matching our sample with the SIMBAD database, the most recent analysis of the MMT hypervelocity star survey \citep{brown14,kreuzer21}, and the list of misclassified hot subdwarfs from the LAMOST survey \citep{luo21}. The main contaminants of our sample are WDs and MS-A/B-type stars as well as a few cataclysmic variables. However, a galaxy and an active galactic nucleus were also previously classified as hot subdwarfs.

\begin{figure*}
  \centering
  \includegraphics[width=\hsize]{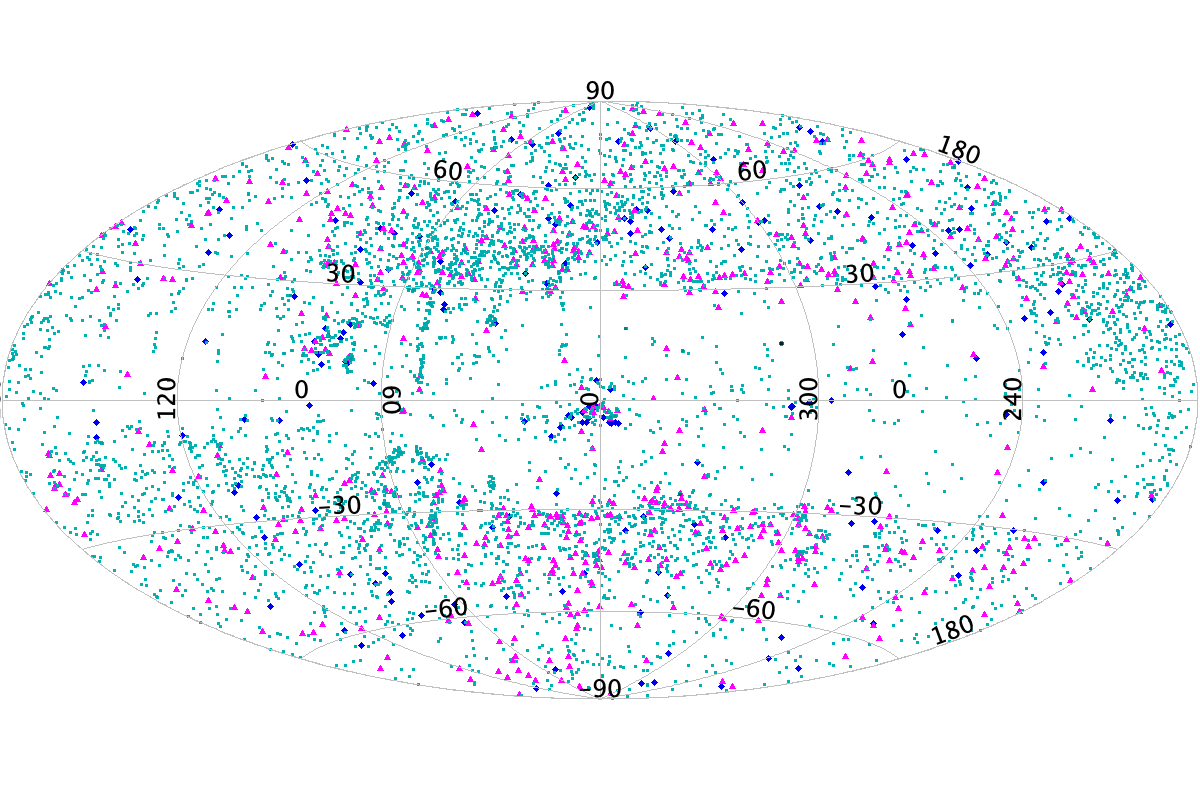}
 \caption{Full-sky distribution of known hot subdwarfs DR3: Known hot subdwarfs with no binarity observed (cyan squares), known hot subdwarfs in wide binary systems (magenta triangles), and known hot subdwarfs in close binary systems with WD, dwarf-M type, or brown dwarf companions (blue diamonds).}
  \label{sky_distn_known_dr3_type}
  \end{figure*}

The 182 misclassified objects are provided with their correct classifications as a separate catalogue. Based on their previous classifications as sdBs or sdOs and their colours, a tentative classification of the WD candidates as either DAB (hydrogen and neutral helium lines) or DAO (hydrogen and/or ionised helium lines) candidates is provided. In the absence of a previous detailed classification, they are classified as WD. The final DR3 catalogue of known hot subdwarfs contains 6616 unique objects. The predominance of hot subdwarfs outside the Galactic plane (see Figure~\ref{sky_distn_known_dr3_type}) is due to observational limitations in crowded regions. For a description of the catalogue columns of both the DR3 catalogue and the catalogue of misclassified objects, see Table~\ref{table:A1}.

 \begin{figure}
  \centering
  \includegraphics[width=\hsize]{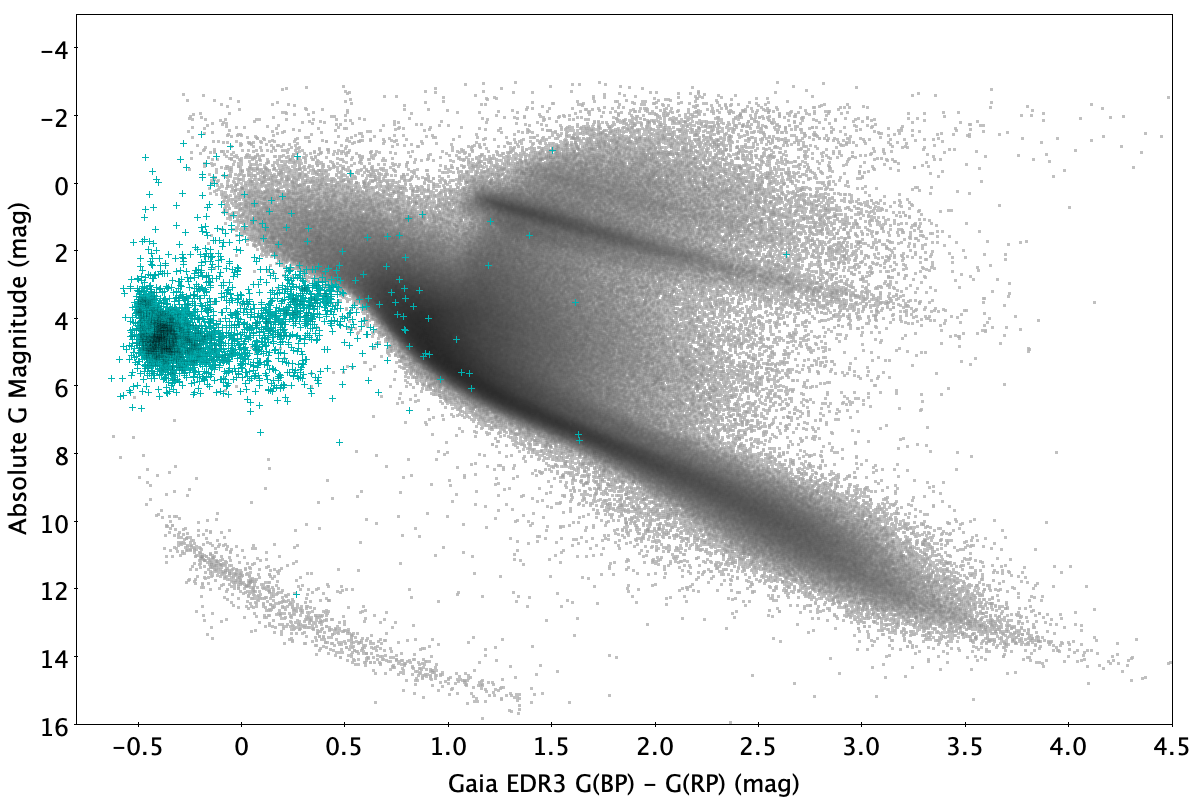}
 \caption{Full-scale {\em Gaia} EDR3 colour--magnitude diagram (grey dots) with the known hot subdwarfs DR3 (cyan crosses). Due to the unreliable absolute G magnitudes of the remaining 2,877 objects, only the 3,373 objects from the DR3 catalogue of known hot subdwarfs that conform to the astrometric and photometric selection quality criteria (see Table 1.) are plotted. Here we can see the EHB cluster of sources at $(G_{\rm BP} - G_{\rm RP})$ -0.35, $G_\textrm{abs}$ 4.5 and the extended cloud of hot subdwarf plus main sequence binaries towards $(G_{\rm BP} - G_{\rm RP})$ 0.5, $G_\textrm{abs}$ 3.0.}
  \label{cmd_gaia_full_known_dr3}
  \end{figure}

\subsection{Classification, spectroscopic parameters, and radial velocities}

For the spectroscopic and photometric classifications, we follow the scheme outlined in \citet{geier17a} and updated in \citet{geier20}. To complement the spectroscopic classes sdB, He-sdB, sdOB, He-sdOB, sdO, and He-sdO, we introduce the additional classes O(H), O(He), PG1159, and $[$WR$]$ to classify the hotter central stars of planetary nebulae (CSPNe) and other post-AGB stars in our sample. O(H) stars have hydrogen-rich spectra similar to sdOs with Balmer lines and usually only one He\,{\sc ii} line at 4686\,{\rm \AA}. O(He) stars exclusively show He\,{\sc ii} and some weaker metal lines of high ionization stages, while PG1159 stars show strong carbon lines in addition to the He\,{\sc ii} lines in the optical. $[$WR$]$ stars are hot and luminous CSPNe with spectra dominated by emission lines.

The catalogue contains spectroscopic parameters such as effective temperatures, surface gravities, and helium abundances. In this catalogue, we aim to provide those parameters for all the stars where such data are provided in the literature. To achieve that in the most efficient way, we started with papers containing parameters of larger samples of hot subdwarfs. For all the remaining stars in the catalogue, we queried the available literature using the SIMBAD database and collected the atmospheric parameters. For the CSPNs, we only include atmospheric parameters determined with non-local thermodynamic equilibrium (NLTE) models, which we consider as reliable (for the references from the literature see Table~\ref{table:A1}).

As the main purpose of this catalogue is the identification and classification of hot subdwarf stars, only one set of atmospheric parameters is provided for each star, even if several different values are provided in the literature. In total, 3087 objects in our catalogue have measured atmospheric parameters taken from the literature. This is, again, a significant increase with respect to the 2187 objects included in \citet{geier20} and should now be close to complete, especially regarding results taken from older literature.

Radial velocities (RVs) are provided for the 2791 stars with spectra in the SDSS and LAMOST data archives. However, for the most helium-rich objects, systematic offsets by up to $\sim100\,{\rm km\,s^{-1}}$ are possible owing to cross-correlation with inadequate template spectra \citep{geier15a,geier17a}. If available, we used the RVs provided by \citet{luo21} for the LAMOST spectra, because they were measured using appropriate models and should not be affected by such systematic errors.

\section{Constructing the hot subluminous star catalogue from {\em Gaia} EDR3}

The aim of the European Space Agency's {\em Gaia} mission was to create a three-dimensional map of our Galaxy. {\em Gaia} Early Data Release 3 (EDR3) \citep{gaia20a} has charted over 1.8 billion sources and contains astrometry and photometry data acquired during the first 34 operational months.
The procedure we used to generate our catalogue is based on that used in \citet{geier19} but modified to reflect the differences between the {\em Gaia} DR2 and EDR3 data sets in accordance with the recommendations made by \citet{fabricius20} that the {\em Gaia} EDR3 data set be considered independently of DR2. As such, we did not directly use any of the results from \citet{geier19}. Furthermore, given the proven high quality of the {\em Gaia} data, we decided to compile the catalogue from these data only and not include ground-based datasets as in \citet{geier19}.

\subsection{Hot subluminous star selection using parallax measurements}

We selected all {\em Gaia} EDR3 objects with good parallax measurements ($\verb!parallax_over_error! > 5$; $\verb!parallax! > 0$), and used the colour criteria ($-0.7 < (G_{\rm BP} - G_{\rm RP}) < 0.7$) and the absolute magnitude criteria ($-5 < G_\textrm{abs} < 7$). The absolute G magnitude was calculated using:

\begin{equation}
\verb!abs_g_mag! = \verb!phot_g_mean_mag! + 5 + 5\log_{10}(\verb!parallax!/1000)
\end{equation}

Applying these criteria resulted in 3,213,406 objects from the {\em Gaia} EDR3 catalogue. The criteria used were less restrictive than the ($-1 < G_\textrm{abs} < 7$; $-0.7 < (G_{\rm BP} - G_{\rm RP}) < 0.7$) used for the \citet{geier19} catalogue. This was done to ensure that all possible hot subluminous stars were considered. Visual comparison of the respective CMDs showed this to be the case. 

We calculated the corrected BP and RP flux excess factor ($\verb!phot_bp_rp_excess_factor_corrected!$) using the equations found in Sect. 6 of \citet{riello20}  and applied the following photometric and astrometric quality criteria: $\verb!astrometric_sigma5d_max! < 1.5$ limiting the five-dimensional uncertainty in the astrometric solution \citep{lindegren18,lindegren21a}; $|\verb!phot_bp_rp_excess_factor_corrected!| < 0.6$ to act as a filter to remove sources with inconsistencies in $G$, $G_{\rm BP}$, and $G_{\rm RP}$ photometry \citep{riello20}.
Application of these quality criteria reduced the number of potential candidates with good parallax to 3,195,369.

The correction for parallax bias with regard to magnitude (zero point correction - zpc) outlined in \citet{lindegren21} was not taken into account for the candidate selection as many of the hot subluminous stars lie in a parameter range where the correction is not well defined. To maintain internal consistency without unnecessarily removing objects without a well-defined zero-point we have opted to use {\em Gaia} EDR3 parallax without bias correction.

It should be noted that the astrometric quality criteria from \citet{lindegren18,lindegren21a} based on the renormalised unit weight error (\texttt{ruwe}) and astrometric excess noise significance (\verb!astrometric_excess_noise_sig!) were not used as they are sensitive to astrometric binaries with unresolved companions and we do not aim to exclude such systems.

\begin{figure}
  \centering
  \includegraphics[width=\hsize]{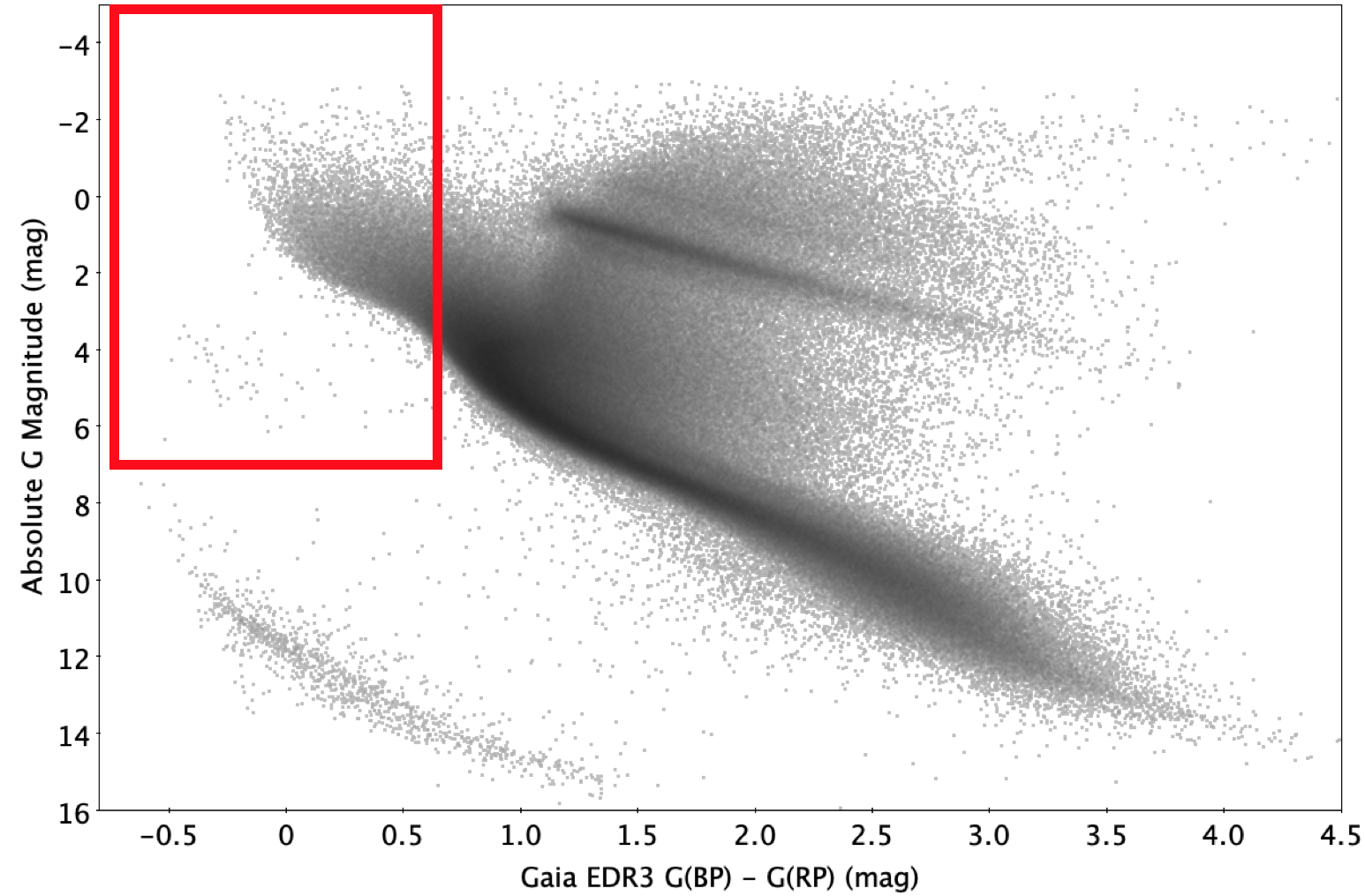}
  \includegraphics[width=\hsize]{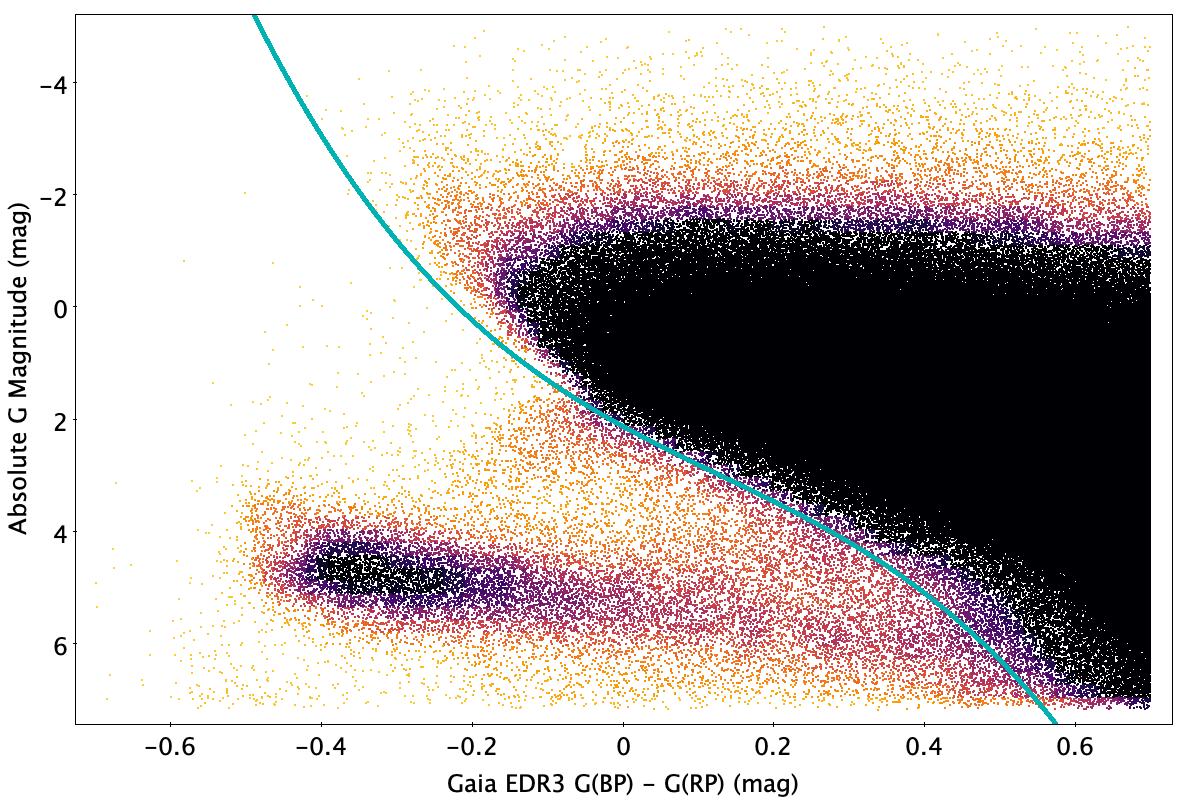}
 \caption{Gaia EDR3 Colour magnitude diagram (CMD). Upper panel: Full-scale {\em Gaia} EDR3 CMD (grey dots) with the {\em Gaia} EDR3 initial hot subluminous star CMD range (red rectangle) used in subsequent CMD plots; lower panel {\em Gaia} EDR3 initial hot subluminous star CMD region. The colour scale shows the number density of objects. The cyan line shows the cutoff used to remove the main sequence region.}
  \label{cmd_numdens}
  \end{figure}

We created a cut in {\em Gaia} EDR3 colour and absolute magnitude parameter space by dividing the {\em Gaia} EDR3 hot subdwarf CMD parameter space into a 100x100 grid and counting the number of objects in each subregion. We generated a polynomial in colour--magnitude space (see Table 1) to follow the trough in values between the hot subdwarf and main sequence clusters of objects and act as a main sequence region rejection criterion (see the cyan line in Figure~\ref{cmd_numdens}).

\begin{table*}[h!]
\centering
\begin{tabular}{c} 
 \hline
 \hline
 \noalign{\smallskip}
 {\it 1. {\em Gaia} EDR3 initial hot subluminous star CMD ranges:} \\
 \noalign{\smallskip}
 \hline
 \noalign{\smallskip}
 $-0.7 \leq (G_{\rm BP} - G_{\rm RP}) \leq 0.7$ \\
 $-5 \leq G_{\rm abs} \leq 7$ \\ [0.5ex]
 \noalign{\smallskip}
 \hline
 \hline
 \noalign{\smallskip}
 \it 2. Astrometric quality selection criteria: \\
 \noalign{\smallskip}
 \hline
 \noalign{\smallskip}
 $\verb!parallax! > 0$ \\ 
 $\verb!parallax_over_error! > 5$ \\ 
 $\verb!astrometric_sigma5d_max! < 1.5$ \\ [0.5ex]
 \noalign{\smallskip}
 \hline
 \hline
 \noalign{\smallskip}
 \it 3. Photometric quality selection criteria: \\
 \noalign{\smallskip}
 \hline
 \noalign{\smallskip}
 $\verb!|phot_bp_rp_excess_factor_corrected|! < 0.6$ \\ [0.5ex]
 \noalign{\smallskip}
 \hline
 \hline
 \noalign{\smallskip}
 {\it 4. {\em Gaia} EDR3 CMD main sequence region rejection criterion:} \\
 \noalign{\smallskip}
 \hline
 \noalign{\smallskip}
 $G_{\rm abs} < 17.7 (G_{\rm BP} - G_{\rm RP})^3 - 6.9 (G_{\rm BP} - G_{\rm RP})^2 + 7.35 (G_{\rm BP} - G_{\rm RP}) + 1.95$ \\ [0.5ex]
 \noalign{\smallskip}
 \hline
 \hline
 \noalign{\smallskip}
 {\it 5. Strict hot subluminous star selection criterion:} \\
 \noalign{\smallskip}
 \hline
 \noalign{\smallskip}
 $G_{\rm abs} < 12.0 (G_{\rm BP} - G_{\rm RP})^3 + 12.9 (G_{\rm BP} - G_{\rm RP})^2 + 6.8 (G_{\rm BP} - G_{\rm RP}) + 3.53$ \\ [0.5ex]
 \noalign{\smallskip}
 \hline
 \hline
\end{tabular}
%\noalign{\smallskip}
\caption{Table of selection criteria applied to {\em Gaia} EDR3 to define the hot subluminous star candidate selection from colour and absolute G magnitude using sources with good parallax.}
\label{table:1}
\end{table*}

Applying the selection criteria for sources with a parallax error of less than 20\% (criteria 1, 2, 3, and 4; see Table~\ref{table:1}) reduced the number of potential hot subluminous stars to 16,959 objects.

It is known that sources in regions of high apparent stellar density are prone to inaccurate astrometry. When these inaccurate measurements are repeatable, the astrometric quality criteria will not flag such measurements as being erroneous \citep{gentile19,gentile21,geier19,pelisoli19,culpan}. Furthermore, regions of high apparent stellar density are also known to be more susceptible to inaccuracies in the determination of the $G_{\rm BP}$ and $G_{\rm RP}$ background as well as blending effects \citep{riello20}.

In order to minimise these crowded region effects, we defined the subset of objects in the {\em Gaia} EDR3 CMD hot subluminous star region with no apparent neighbours within a 5 arcsec radius. We find 10,672 such candidates and refer to them here as the Parallax Selection 1 set. We next considered those objects with one or more apparent neighbours within 5 arcsec. The total G flux from the hot subluminous star and the apparent neighbours within 5 arcsec was calculated. Next we selected the hot subluminous stars whose G flux dominated the 5 arcsec neighbourhood flux. The criterion for this was at least 70\% of the 5 arcsec neighbourhood G band flux coming was from the hot subluminous star itself. There are 1,147 such candidates, which we refer to as the Parallax Selection 2 set. 

\begin{figure*}
  \centering
  \includegraphics[width=\hsize]{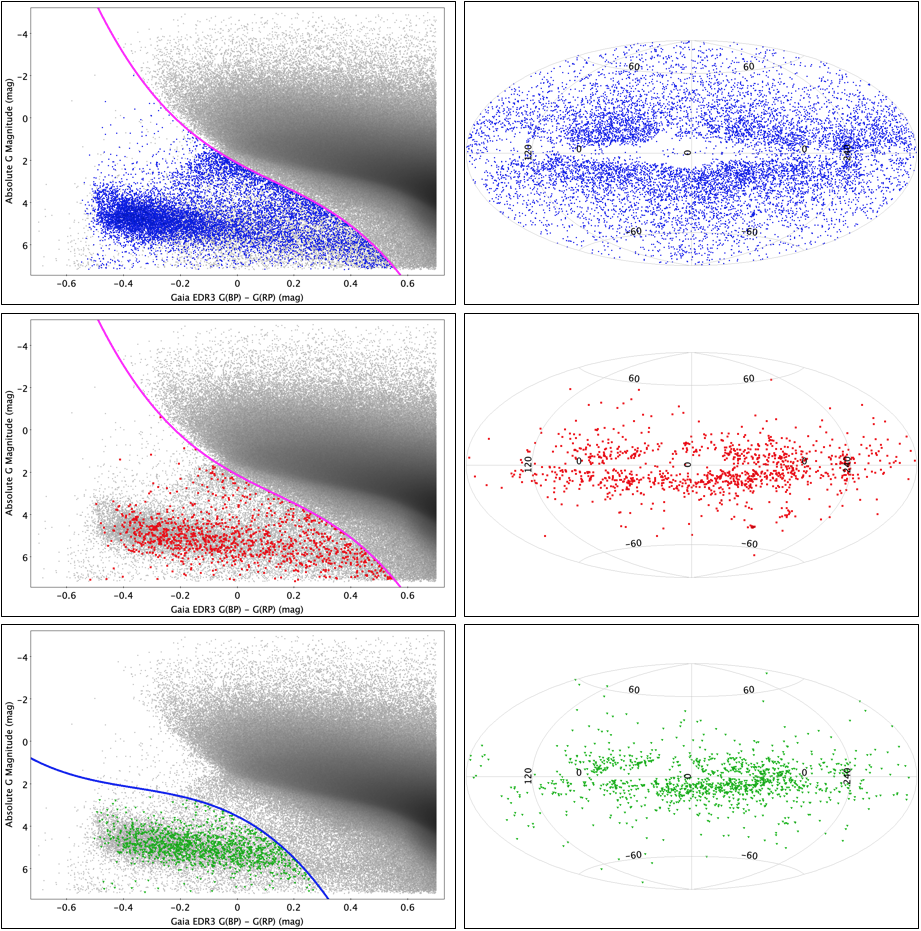}
 \caption{Left column: {\em Gaia} EDR3 colour--magnitude diagrams showing the 3,213,406 initial candidate objects (grey dots) with the {\em Gaia} EDR3 CMD hot subluminous star region selection criteria (magenta line - top and middle rows) and the strict hot subluminous star-selection criterion (blue line - bottom row) (see Table~\ref{table:1}). Right column: Sky distribution of the parallax selection candidate objects. Top row: Parallax Selection 1 objects (blue circles). Middle row: Parallax Selection 2 objects (red squares). Bottom row: Parallax Selection 3 objects (green triangles).}
  \label{cmd_sky_1_2_3}
  \end{figure*}

Finally, we considered the remaining candidate objects whose G flux was less than 70\% of the total 5 arcsec neighbourhood G flux. As both the astrometry and photometry might be adversely affected for these objects, we applied a stricter parallax quality criterion (parallax error < 10\%) and a stricter cutoff (see criterion 5 in Table~\ref{table:1}) to ensure that only the objects whose colour lies directly within the most populated region of the hot subluminous star CMD cloud were selected. It was considered highly unlikely that objects that were bluer than a hot subdwarf were then reddened by extinction to make them lie in this region. We generated a polynomial for the strict hot subdwarf selection criterion (see Figure~\ref{cmd_sky_1_2_3} bottom left panel). There are 1,304 candidate objects in this final Parallax Selection 3 set.

The three parallax selection sets identified here (see Figure~\ref{cmd_sky_1_2_3}) resulted in 13,123 hot subluminous stars (Final Parallax Selection) selected using colour and absolute G magnitude. A summary of the selection criteria used and their effect on the number of candidate objects found in the {\em Gaia} EDR3 hot subluminous star catalogue can be seen in Table~\ref{table:3}.

\subsection{Hot subluminous star parameter determination for candidate selection using proper motion measurements}\label{redPM_parameters}

The use of reduced proper motion as a proxy for absolute G magnitude when identifying hot subluminous stars, white dwarfs, and BHB objects has become common practice, and was used by \citet{gentile21}, \citet{geier19}, and \citet{culpan}.

Using parallax errors <20\% as a selection criterion effectively limits the distance at which we are able to select hot subluminous stars. Around 90\% of the Final Parallax Selection is found at distances of less than 3 kpc from the Sun. By using reduced proper motion as a proxy for absolute G magnitude, we are able to benefit from the fact that there are many objects in the {\em Gaia} EDR3 data set that have a parallax error >20\% yet still have reliable proper motion measurements.

\begin{figure}
  \centering
  \includegraphics[width=\hsize]{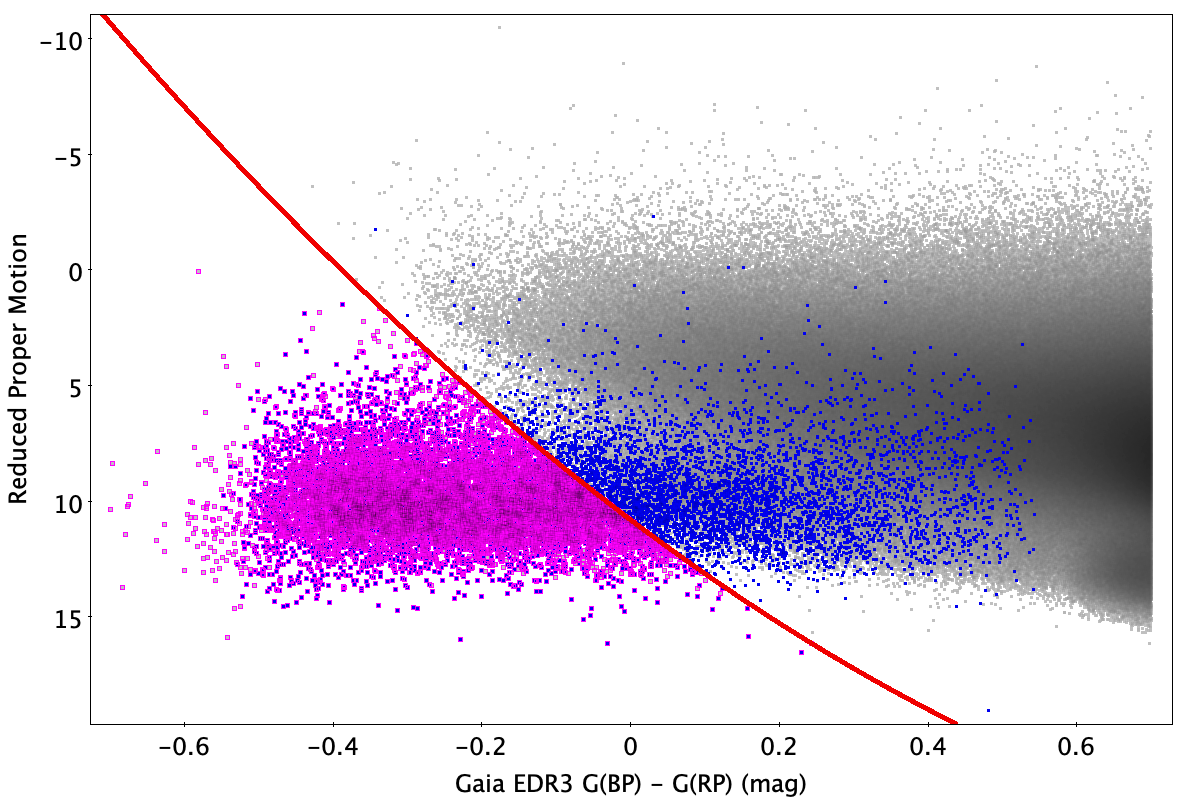}
  \includegraphics[width=\hsize]{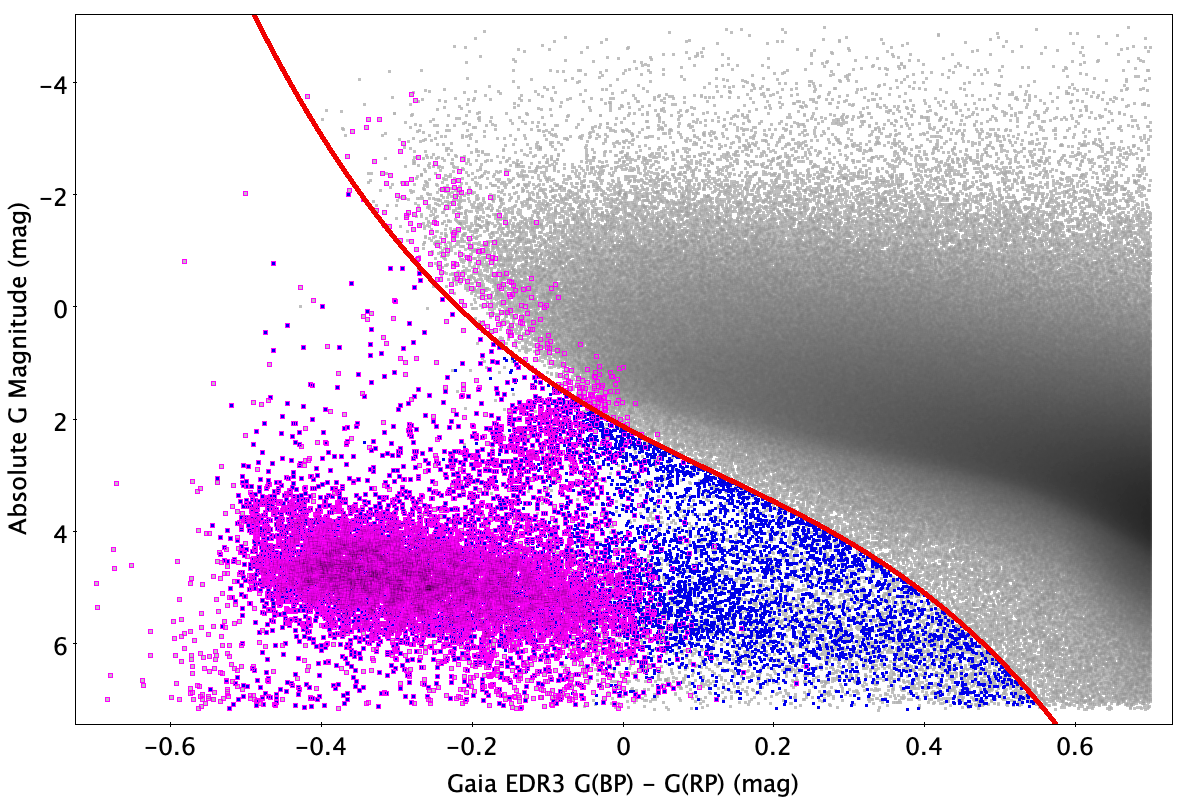}
 \caption{Upper panel: {\em Gaia} EDR3 colour-reduced proper motion diagram. Lower panel: {\em Gaia} EDR3 colour--magnitude diagram. Both showing the initial parallax selection (grey dots) and Final Parallax Selection (blue circles) and reduced proper motion selection (magenta squares). The polynomial lines are defined as the cutoff criteria to remove the main sequence objects in colour-reduced-proper-motion space and colour--magnitude space, respectively.}
  \label{crpm_cmd_parallax_redpm}
  \end{figure}

Plotting the Final Parallax Selection objects in colour reduced proper motion space (see Figure~\ref{crpm_cmd_parallax_redpm} upper panel) we were able to find the parameter space where we could expect to discover more distant candidate objects. In order to do this, the $\verb!reduced_proper_motion!$ ($H_{\rm G}$) and the $\verb!proper_motion_over_error!$ ($\verb!pm!/\sigma_{\rm pm}$) were calculated for all 3,213,406 objects in the {\em Gaia} EDR3 hot subluminous star CMD range where:

\begin{equation}
    H_G = G + 5\log_{10}(\mu) + 5
\end{equation}
\begin{equation}
    \sigma_{pm} = \frac{\sqrt{(\sigma_{pmra}\texttt{pmra})^2 + (\sigma_{pmdec}\texttt{pmdec})^2}}{pm}
\end{equation}

The plot of colour versus reduced proper motion for the 3,213,406 initial selection and the Final Parallax Selection superimposed (see Figure~\ref{crpm_cmd_parallax_redpm}, upper panel) shows, as expected, the same general form as the colour--magnitude plot (lower panel), but with the hot subluminous star region less well separated from the main sequence region. This gave us confidence that using reduced proper motion as a proxy for absolute magnitude is a valid method for this data set albeit with lower completeness and higher levels of contamination than found in the Final Parallax Selection. A discussion of the relative contamination and completeness between the Final Parallax Selection and the Proper Motion Selection can be found in Section 5.

As was done in the parallax selection procedure, a polynomial line was defined to separate the cloud of hot subluminous stars and the main sequence objects. For the reduced proper motion selection, this was done iteratively by comparing the results for the same objects in the colour versus reduced proper motion (colour-$H_{\rm G}$) diagram (see Figure~\ref{crpm_cmd_parallax_redpm} upper panel) and the CMD (see Figure~\ref{crpm_cmd_parallax_redpm} lower panel).

\subsection{Reduced proper motion hot subdwarf candidate selection}

We found 7,390,541 objects within the hot subluminous star colour--$H_{\rm G}$ region given in section 3.3 (see Table \ref{table:2}). The use of proper motion (\verb!pm!) in calculating reduced proper motion is directly analogous to the use of parallax in calculating absolute magnitude. Thus, an equivalent quality criterion of proper motion error < 20\% was applied \citep{gentile21,culpan}. This reduces the dataset to 6,860,074 objects.

Comparison of the colour--$H_{\rm G}$ plot for the parallax error < 20\% objects (see Figure~\ref{crpm_cmd_parallax_redpm} upper panel) and the parallax error $\geq$20\% objects (see Figure~\ref{cmd_crpm_redpm_numdens}) showed that a very different population of stars is present in the parallax error $\geq$20\% selection. Using a parallax error <20\% cutoff criterion is, as stated in section \ref{redPM_parameters}, effectively limiting the distance of the selected objects to 2-3 kpc from the Sun. Considering objects at greater distances brings with it considerable contamination from the Magellanic Clouds. The region around the Magellanic Clouds was therefore removed, reducing further the number of objects to 4,856,823.
Applying the astrometric and photometric quality criteria and the {\em Gaia} EDR3 colour--$H_{\rm G}$ diagram main sequence region rejection criterion that was found in Section \ref{redPM_parameters} (see Table 2.) to the remaining sources leaves 92,409 hot subluminous candidates.

We then applied the same filtering criteria for crowded region effects as used in the Parallax Selection 1 and Parallax Selection 2, resulting in 66,393 objects with no apparent neighbours within 5 arcsec (Proper Motion Selection 1) and 7,044 objects where at least 70\% of the total flux from within a 5 arcsec radius came from the candidate object (Proper Motion Selection 2).

\subsection{Proper motion selection: white dwarf contamination}

\begin{figure}
  \centering
  \includegraphics[width=\hsize]{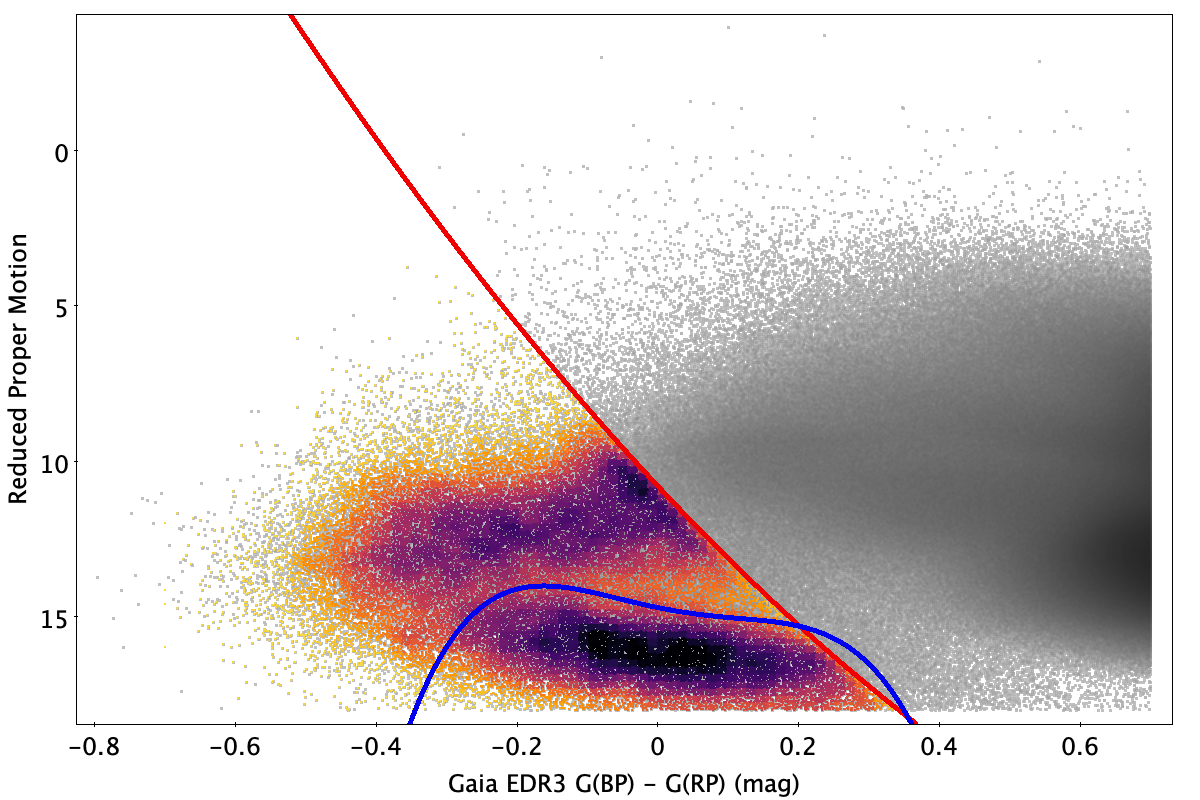}
 \caption{{\em Gaia} EDR3 colour-reduced proper motion diagram for the hot subluminous star region. The grey dots show the objects outside the Magellanic Clouds that conform with all quality criteria. The colour scale shows the number density of Proper Motion Selection 1 and 2 objects in colour--$H_{\rm G}$ space. The red line shows the cutoff used to remove the main sequence region. The blue line shows the cutoff used to remove the white dwarf region.}
  \label{cmd_crpm_redpm_numdens}
  \end{figure}

We saw a cloud of objects that were present at high reduced proper motions (see Figure~\ref{cmd_crpm_redpm_numdens}). These were not seen in the Final Parallax Selection (see Figure~\ref{crpm_cmd_parallax_redpm} upper panel). We made a colour--$H_{\rm G}$ plot with the number density plotted on the colour axis using the same method as outlined in Section 3.1 and observed two clear concentrations of objects in colour--$H_{\rm G}$ space. Overlaying the high-probability white dwarf candidates in the reduced proper motion extension from \citet{gentile21} shows that the cloud at higher reduced proper motions is white dwarf contamination. An additional cutoff was therefore made to reduce this white dwarf contamination (see Figure~\ref{cmd_crpm_redpm_numdens}). Applying the white-dwarf-rejection criterion leaves 48,462 hot subluminous star candidates (Final Proper Motion Selection). For a summary of the selection criteria and their effects on the number of candidate objects, see again Table~\ref{table:3}.

\begin{table*}[h!]
\centering
\begin{tabular}{c} 
 \hline
 \hline
 \noalign{\smallskip}
 {\it 1. {\em Gaia} EDR3 hot subluminous star colour-$H_{\rm G}$ ranges:} \\
 \noalign{\smallskip}
 \hline
 \noalign{\smallskip}
 $-0.7 \leq (G_{\rm BP} - G_{\rm RP}) \leq 0.7$ \\
 $-10 \leq H \leq 17$ \\ [0.5ex]
 \noalign{\smallskip}
 \hline
 \hline
 \noalign{\smallskip}
 \it 2. Astrometric quality selection criteria: \\
 \noalign{\smallskip}
 \hline
 \noalign{\smallskip}
 $\verb!parallax_over_error! \leq 5$ \\ 
 $\verb!pm_over_error! > 5$ \\ 
 $\verb!astrometric_sigma5d_max! < 1.5$ \\ [0.5ex]
 \noalign{\smallskip}
 \hline
 \hline
 \noalign{\smallskip}
 \it 3. Photometric quality selection criteria: \\
 \noalign{\smallskip}
 \hline
 \noalign{\smallskip}
 ($\verb!phot_bp_n_obs! > 2$ AND  $\verb!phot_rp_n_obs! > 2$)
  OR $\verb!phot_g_mean_mag! < 19$\\ [0.5ex]
 \noalign{\smallskip}
 \hline
 \hline
 \noalign{\smallskip}
 {\it 4. {\em Gaia} EDR3 colour-H\_G main sequence region rejection criterion:} \\
 \noalign{\smallskip}
 \hline
 \noalign{\smallskip}
 $H < -9.26 (G_{\rm BP} - G_{\rm RP})^2 + 24.4 (G_{\rm BP} - G_{\rm RP}) + 10.8$ \\ [0.5ex]
 \noalign{\smallskip}
 \hline
 \hline
 \noalign{\smallskip}
 {\it 5. White dwarf region rejection criterion:} \\
 \noalign{\smallskip}
 \hline
 \noalign{\smallskip}
 $H < 342.5 (G_{\rm BP} - G_{\rm RP})^4 + 40.8 (G_{\rm BP} - G_{\rm RP})^3 + 13.7 (G_{\rm BP} - G_{\rm RP})^2 + 4.6 (G_{\rm BP} - G_{\rm RP}) + 14.7$ \\ [0.5ex]
 \noalign{\smallskip}
 \hline
 \hline
 \noalign{\smallskip}
 {\it 6. Magellanic Cloud Rejection Criteria:} \\
 \noalign{\smallskip}
 \hline
 \noalign{\smallskip}
 within 15$^\circ$ of RA = 81.28$^\circ$, DEC = -69.78$^\circ$ OR
 within 9$^\circ$ of RA = 12.8$^\circ$, DEC = -73.15$^\circ$  \\ [0.5ex]
 \noalign{\smallskip}
 \hline
 \hline
\end{tabular}
%\noalign{\smallskip}
\caption{Table of selection criteria applied to {\em Gaia} EDR3 to define the hot subdwarf candidate selection from colour and reduced proper motion using sources with unreliable parallax measurements}
\label{table:2}
\end{table*}

\section{Indications of variability}

The formation and evolution of hot subdwarf stars is still not fully understood but mass loss and mass transfer with binary partners is thought to play a major role \citep{geier17a,pelisoli20}. We examined both the astrometric and photometric variability using different methods when applied to the {\em Gaia} EDR3 hot subluminous star catalogue and the known hot subdwarf catalogue and compared the results.

Astrometric variability, where the centre of an object moves between different images, is a known method to detect stellar companions \citep{lindegren18,lindegren21a,belokurov20,penoyre21}. The use of photometry to detect variability faces additional challenges as the uncertainty in the mean flux published in {\em Gaia} EDR3 is the standard deviation of the flux measurements with a weighting factor applied. However, this weighting factor is not provided. Many of the published methods \citep{chornay21,mowlavi21} assume that the weighting factors are equal. The errors in the photometric noise are not only dependent on the variability of a candidate object, but also on colour, magnitude, sky background, and stray light. Different apparent magnitudes also have different acquisition and calibration parameters. Despite these limitations, the consideration of sources with anomalous flux error has led to successful identification of variable objects \citep{gentile21,chornay21,mowlavi21,guidry21}.

\subsection{Photometric variability}

To investigate the photometric variability, we applied the methods described in \citet{gentile21} and \citet{guidry21} and compared the results. Both of these methods were developed to identify photometric variability in white dwarfs which are close neighbours to hot subdwarfs in the Hertzsprung-Russell diagram. As such we considered it likely that these methods could be applied to hot subdwarfs.

\begin{figure}
  \centering
  \includegraphics[width=\hsize]{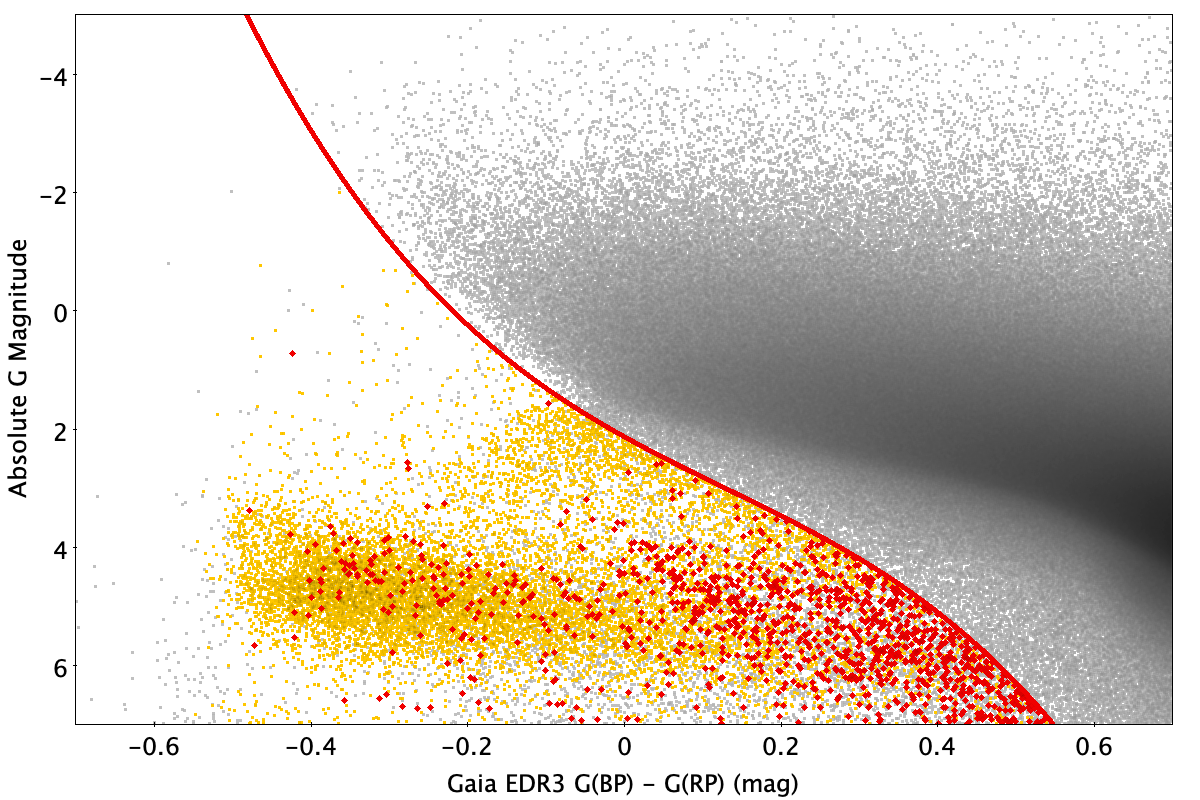}
 \caption{{\em Gaia} EDR3 colour--magnitude diagram for the hot subluminous star region showing the initial parallax selection (grey dots), Final Parallax Selection (yellow circles), and hot subluminous stars found to have an excess flux error that indicates variability. The red line shows the cutoff used to remove the main sequence region.}
  \label{cmd_variable_final_parallax}
  \end{figure}

The excess flux error method presented in \citet{gentile21} compares the $\verb!phot_g_flux_error!$ of each candidate object to the median G flux error of 500 similar objects in terms of colour ($\verb!bp_rp!$), G flux ($\log_{10}\verb!phot_g_mean_flux!$), and the number of observations ($\verb!phot_g_n_obs!$) taken from the full {\em Gaia} EDR3 catalogue. A more precise description of the method and the evaluation can be found in \citet{gentile21}. This method, when applied to the Final Parallax Selection, found that 9.6\% (1,308 of 13,123) of the candidate objects with a parallax error < 20\% showed an excess of flux error indicating variability (see Figure~\ref{cmd_variable_final_parallax}).

\begin{figure}
  \centering
  \includegraphics[width=\hsize]{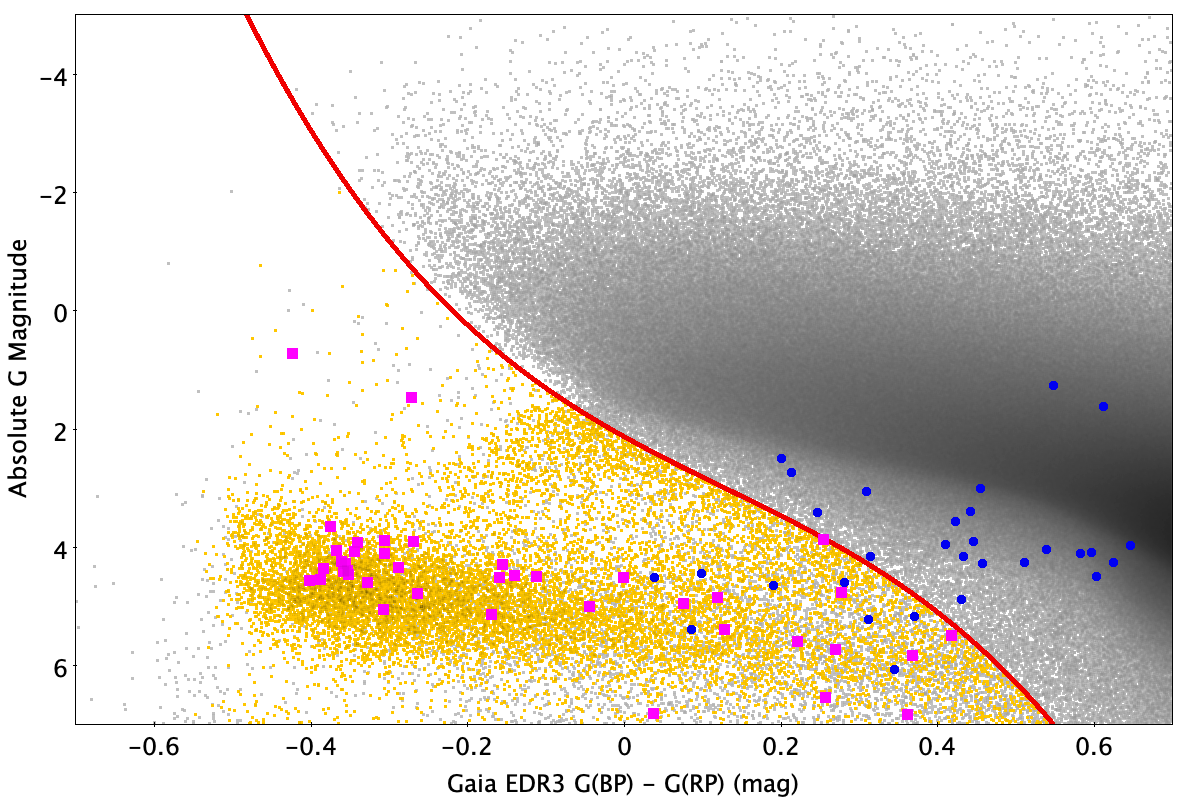}
 \caption{{\em Gaia} EDR3 colour--magnitude diagram for the hot subluminous star region showing the initial parallax selection (grey dots), Final Parallax Selection (yellow circles), known hot subdwarfs in wide binary systems displaying an excess in flux error (blue circles), and known hot subdwarfs in close binary systems that display an excess in flux error (magenta squares).The red lines show the variability cutoffs used.}
  \label{cmd_variable_known_binaries}
  \end{figure}

This method was also applied to the catalogue of known hot subdwarfs. We found that 7.6\% of hot subdwarfs in wide binary systems and 23\% of hot subdwarfs in close binary systems (see Figure~\ref{cmd_variable_known_binaries}) showed an excess in flux error. Also, 5.2\% of hot subdwarfs with no indication of binarity display an excess in flux error.

The results show a higher incidence of variability in composite systems when compared to single star systems. This might be due to intrinsic variability of the cool main sequence companions, which tend to show variations caused by rotation and spots \citep{pelisoli20}. The percentages found also show that close binary systems are more likely to exhibit photometric variability that is detectable with this method than wide binary systems. This is unsurprising as the amplitude of photometric variations is larger in close binary systems when compared to wider binaries with main sequence companions. Furthermore, there is a significant fraction of close binary systems that show eclipses, reflection effects or ellipsoidal variations among the hot subdwarf population \citep[e.g.][]{schaffenroth19,pelisoli21}.

The method to detect variability that was proposed by \citet{guidry21} calculates a {\em Gaia} variability metric using {\em Gaia} photometry (G flux error, apparent G magnitude, G flux, and the number of observations) and a Zwicky variability metric using photometry data from the Zwicky Transient Facility. This method is calibrated to white dwarfs.

The {\em Gaia} variability metric, when used alone, was found to be less effective than the excess flux error method for our data set. We considered this to be due to the fact that this was only one part of the full method and did not include colour as a comparison criterion. Comparing the results of the two methods revealed an unwanted colour dependency in the levels of variability found using the {\em Gaia} variability index from \citet{guidry21}.

\subsection{Astrometric variability}

\begin{figure}
  \centering
  \includegraphics[width=\hsize]{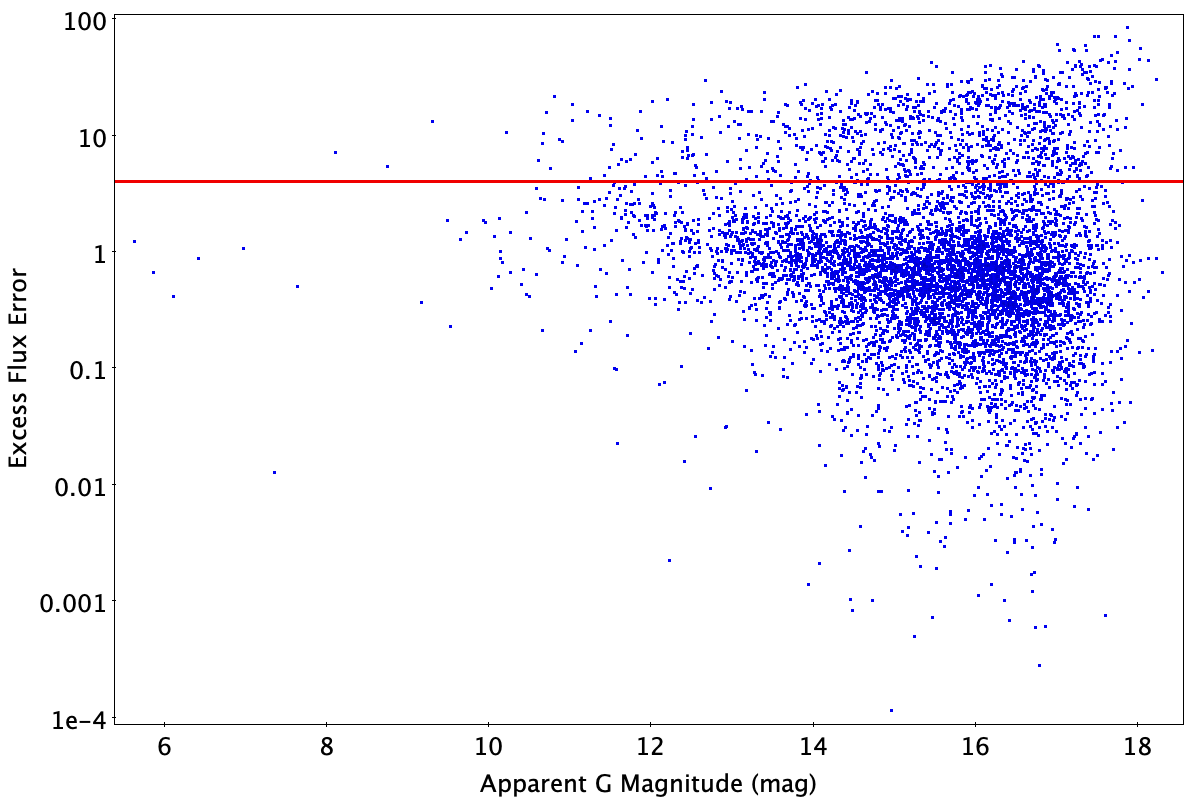}
  \includegraphics[width=\hsize]{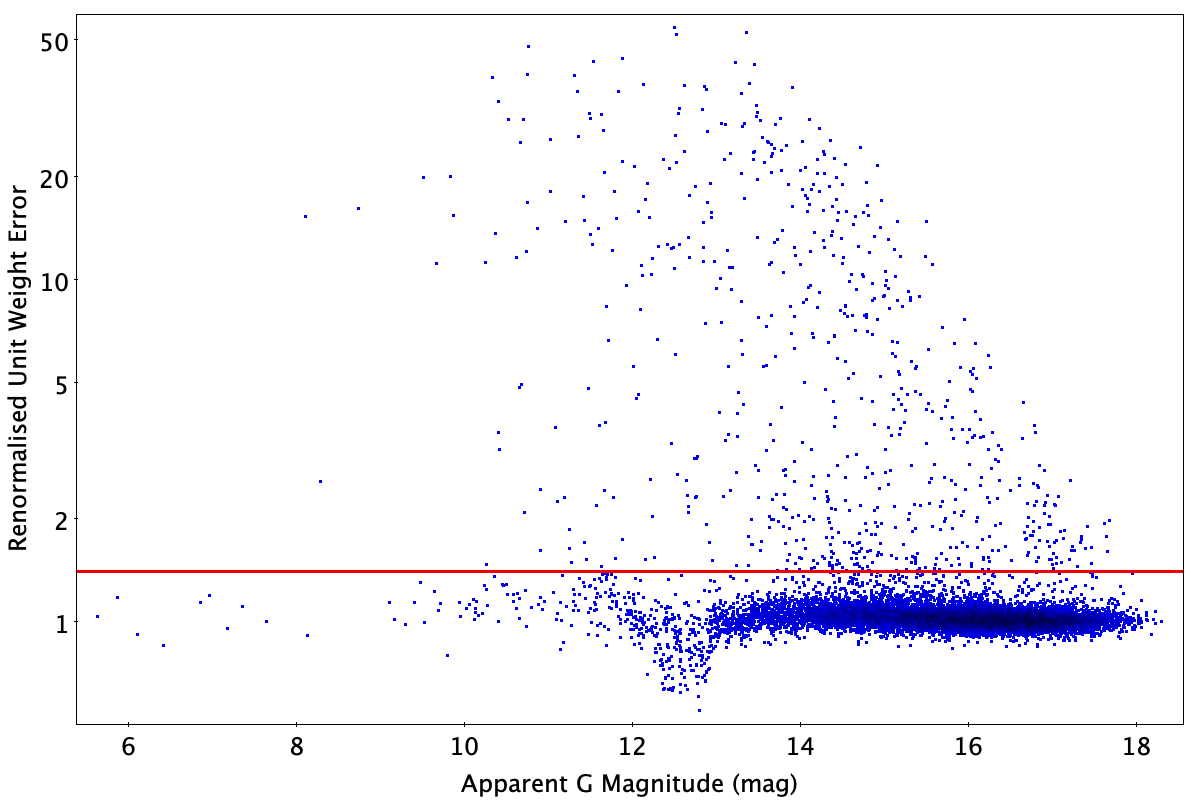}
 \caption{Astrometric and photometric fluctuation indicators versus absolute G magnitude. Upper panel: Distribution of excess flux error versus apparent G magnitude for all sources in the final parallax selection. The horizontal red line shows the cutoff used as the limit for variability detection. Lower panel: Distribution of re-normalised unit weight error versus apparent G magnitude for all sources in the final parallax selection. The horizontal red line shows the cutoff used as the limit for variability detection.}
  \label{variability_versus_appmag}
  \end{figure}

\begin{figure*}
  \centering
  \includegraphics[width=\hsize]{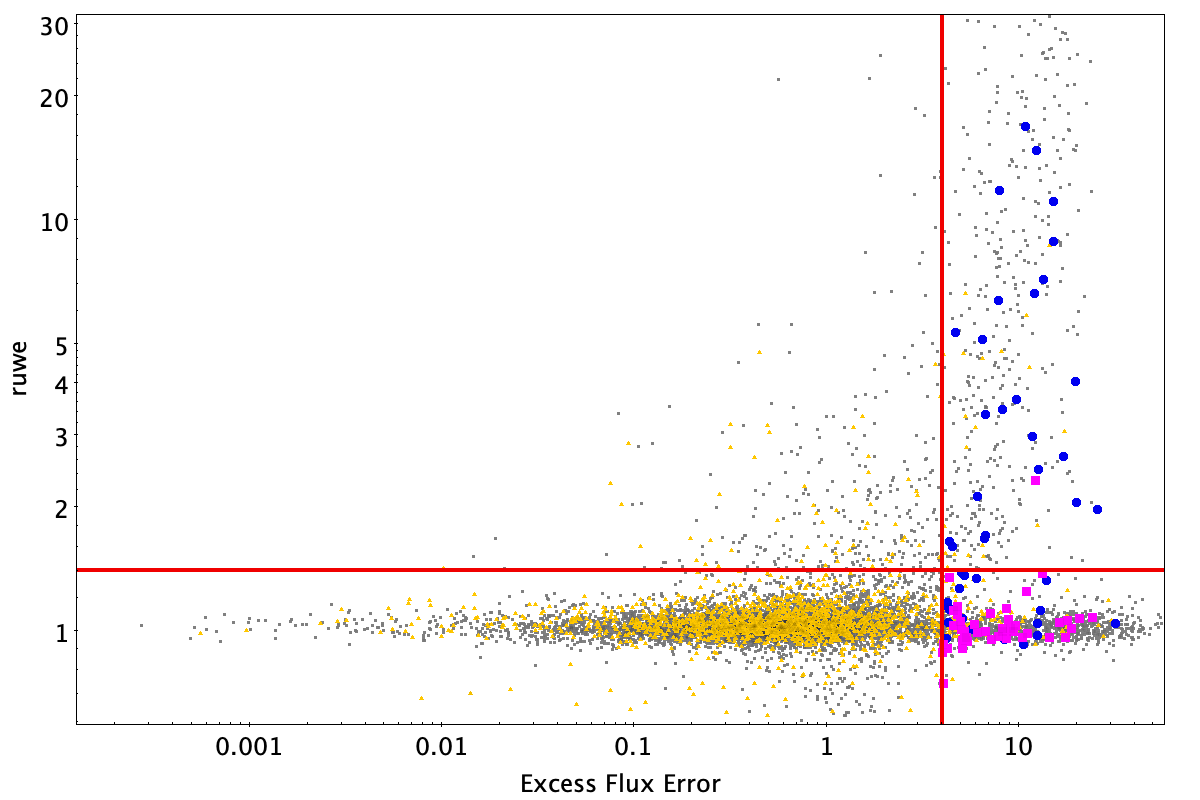}
 \caption{Excess flux error versus {\em Gaia} EDR3 (\texttt{ruwe}) diagram showing the Final Parallax Selection (grey dots), the known hot subdwarfs in wide binary systems (blue circles), the known hot subdwarfs in close binary systems with WD, dM type, or BD companions (magenta squares), and the known hot subdwarfs where no binarity has been observed (yellow triangles). The red lines shows the variability cutoffs used for (\texttt{ruwe}) and flux error excess.}
  \label{ruwe_excess_flux_error}
  \end{figure*}

We used the {\em Gaia} EDR3 output renormalised unit weight error ($\verb!ruwe!$) \citep{lindegren18,lindegren21a} ---a dimensionless measure of how much the centre of an object moves between different images--- as an indicator for possible binarity. Identifying binarity from astrometric variability is currently planned for {\em Gaia} Data Release 3. Plotting $\verb!ruwe!$ versus the excess flux error (see Figure~\ref{ruwe_excess_flux_error}) shows that there is a correlation between photometric and astrometric error excesses. Taking $\verb!flux_error_excess!$ > 4 \citep{gentile21} (see Figure~\ref{variability_versus_appmag} upper panel) and $\verb!ruwe!$ > 1.4 (see Figure~\ref{variability_versus_appmag} lower panel) as the regions where variability has been detected, we see that known hot subdwarfs in wide binary systems display excess errors in both the astrometric and photometric domains. Hot subdwarfs with close binary partners display only photometric error excesses.

As the known composite hot subdwarf binaries with solved orbits have periods of the order of years \citep[e.g.][]{vos18}, they should also be detectable by {\em Gaia} astrometry due to the comparable duration of the {\em Gaia} mission. It is therefore not surprising that many of them show astrometric excesses in addition to the photometric variability discussed in the previous section. The known hot subdwarfs in close binary systems have orbital periods of the order of hours to days \citep[e.g.][]{kupfer15} and are not expected to show astrometric variability.

\begin{figure*}
  \centering
  \includegraphics[width=\hsize]{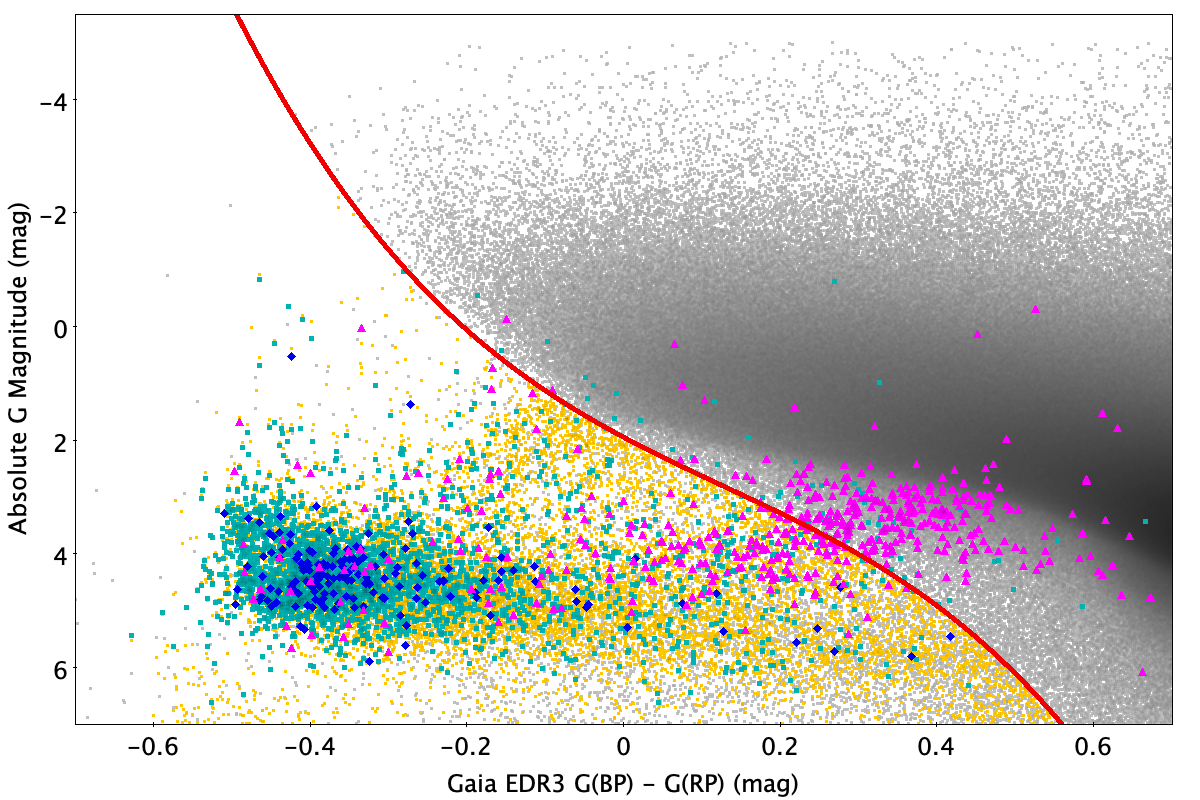}
 \caption{{\em Gaia} EDR3 colour--magnitude diagram for the hot subluminous star region showing the initial parallax selection (grey dots), Final Parallax Selection (yellow circles), known hot subdwarfs with no binarity observed (cyan squares), known hot subdwarfs in wide binary systems (magenta triangles), and known hot subdwarfs in close binary systems with WD, dwarf-M type or brown dwarf companions (blue diamonds).The red line shows the cutoff used to remove the main sequence region.}
  \label{cmd_parallax_final_geier}
  \end{figure*}

\begin{figure}
  \centering
  \includegraphics[width=\hsize]{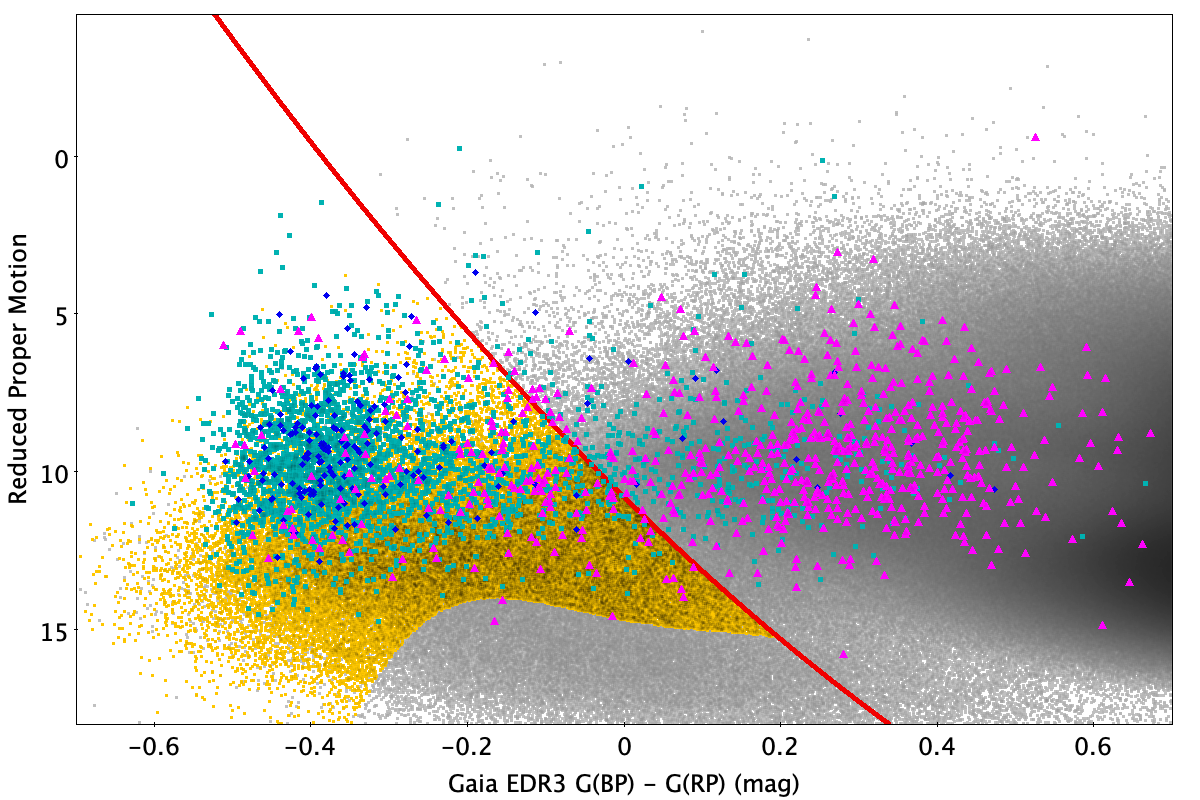}
 \caption{{\em Gaia} EDR3 colour reduced proper motion diagram for the hot subluminous star region showing the initial proper motion selection (grey dots), Final Proper Motion Selection (yellow circles), known hot subdwarfs with no binarity observed (cyan squares), known hot subdwarfs in wide binary systems (magenta triangles), and known hot subdwarfs in close binary systems with WD, dwarf-M type or brown dwarf companions (blue diamonds).The red line shows the cutoff used to remove the main sequence region.}
  \label{crpm_pm_final_geier}
  \end{figure}

\begin{figure}
  \centering
  \includegraphics[width=\hsize]{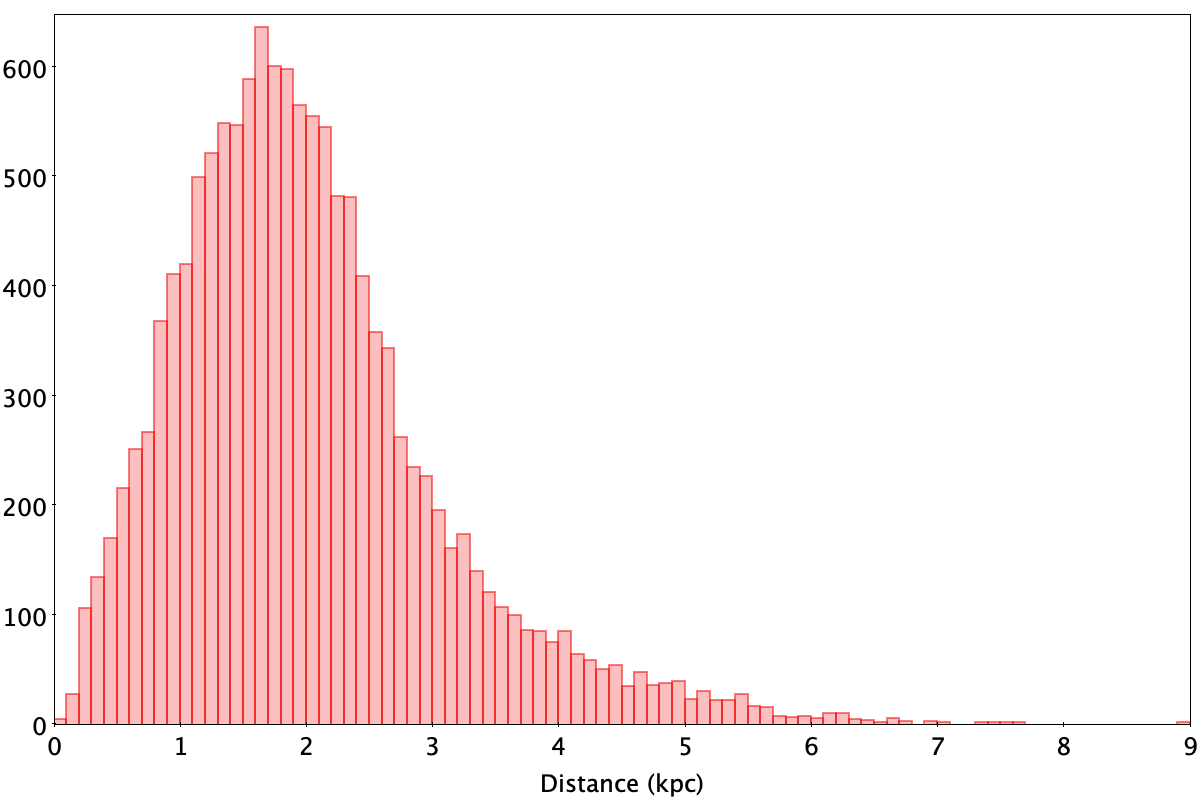}
 \caption{Distance distribution of the {\em Gaia} EDR3 Final Parallax Selection objects.}
  \label{final_parallax_distance_distn}
  \end{figure}

\begin{figure}
  \centering
  \includegraphics[width=\hsize]{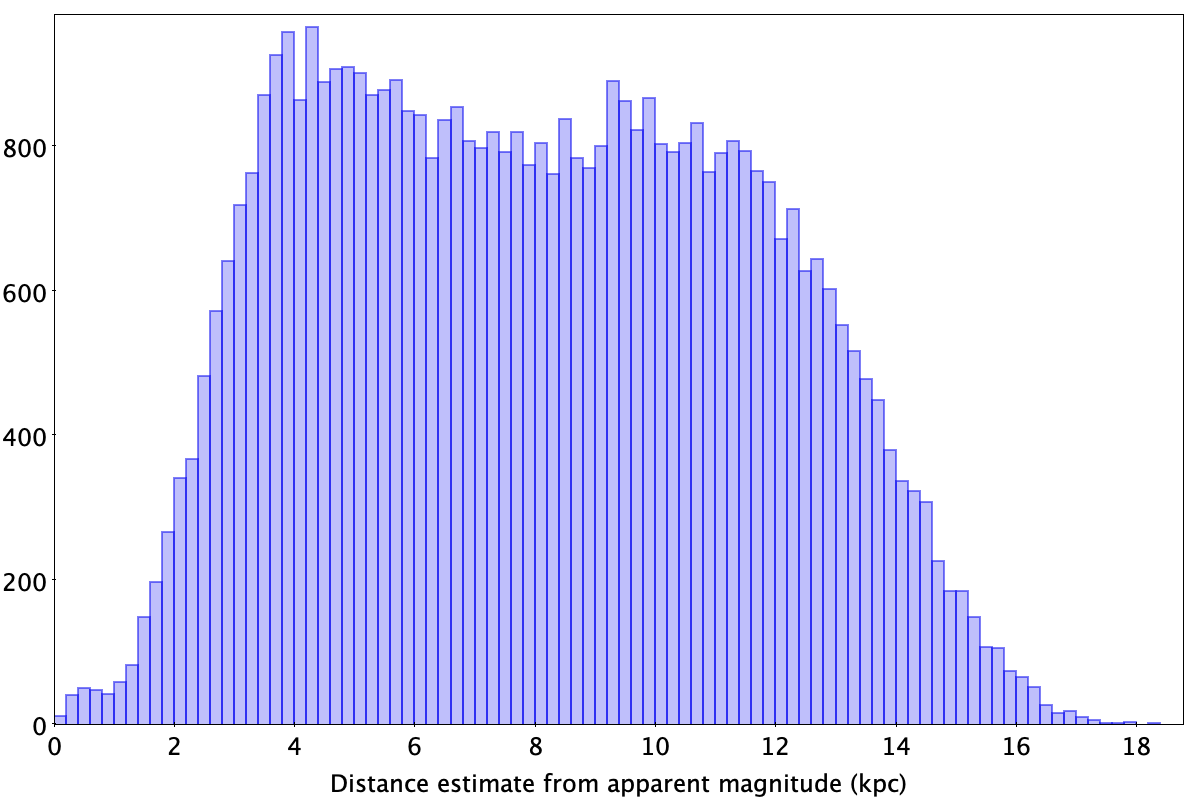}
 \caption{Approximate distance distribution of the {\em Gaia} EDR3 Final Proper Motion Selection objects.}
  \label{final_redPM_distance_distn}
  \end{figure}

\section{Completeness and contamination}

To estimate the completeness and contamination level of the Gaia EDR3 catalogue of hot subluminous stars, we made some comparisons within the catalogue itself and also with the catalogue of known hot subdwarfs.

\subsection{Parallax versus proper motion selection}

When considering all objects with a parallax error $\leq20\%$ we find 8,694 of the 13,123 objects in the Final Parallax Selection are also selected using the proper motion selection criteria. This gives us a $66\%$ completeness of the proper motion selection compared to the parallax selection. 

There is a strong colour dependence in the completeness of the proper motion selection versus the parallax selection (see Figure~\ref{crpm_cmd_parallax_redpm} lower panel). The reduced proper motion selection criteria find nearly all of the Final Parallax selection in the range $-0.7 < G_{\rm BP} - G_{\rm RP} < -0.15$ but find virtually no redder objects where $G_{\rm BP} - G_{\rm RP} \geq -0.15$.

Furthermore, we find only 10,512 parallax error $\leq 20\%$ objects were selected using the proper motion criteria of which 2,967 were not in the Final Parallax Selection. This gives a 28\% contamination of the proper motion selection compared to the parallax selection.

The peak in the distance histogram (see Figure~\ref{final_parallax_distance_distn}) at 1.8\,kpc indicates that the Final Parallax Selection is reasonably complete out to this distance for those hot subdwarfs that are not in binary systems or have unseen binary companions. When we make the approximation that all hot subdwarfs have an absolute magnitude of 4.5, then we can calculate an approximate distance to the  objects selected according to their reduced proper motion using:

\begin{equation}
    d \approx 10^{(0.2 (\verb!phot_g_mean_mag! - 4.5 + 5))}
\end{equation}

Using this equation, we find that the distance distribution of the Final Proper Motion Selection (see Figure~\ref{final_redPM_distance_distn}) peaks at $\sim$4 kpc with candidates found out to $\sim$9 kpc.

\subsection{Known hot subdwarf and {\em Gaia} EDR3 subluminous star catalogue comparison}

We find 3,847 of the known hot subdwarfs to have a parallax error of <20\%, of which 3,246 (84\%) are found within the Final Parallax Selection. As can be seen in Figure~\ref{cmd_parallax_final_geier}, the majority of the known hot subdwarfs that are not in the Final Parallax Selection are those in binary systems with main sequence companions which, due to their redder composite colour, are located beyond our colour cut. 

Of the remaining 2,759 known hot subdwarfs that do not have a parallax error of <20\%, we find 1,715 (62\%) in the Final Proper Motion Selection (see Figure~\ref{crpm_pm_final_geier}). In total, 5,562 (84\%) of the 6,616 known hot subdwarfs are also present in the {\em Gaia} EDR3 hot subluminous star catalogue.

In summary, the {\em Gaia} EDR3 hot subluminous star catalogue parallax selection should be almost complete (80\%-90\%) for single sources and unresolved binaries. However, there is  a much lower completeness for hot subdwarfs in wide binary systems as the main sequence companion moves the colour of the binary system away from the EHB cloud in colour--magnitude space. The reduced proper motion selection has a completeness of~50\% with a contamination level of~28\%.

%--------------------------------------------------------------------
\section{Magnitude, distance, and sky coverage of the catalogues}

\begin{figure*}
  \centering
  \includegraphics[width=\hsize]{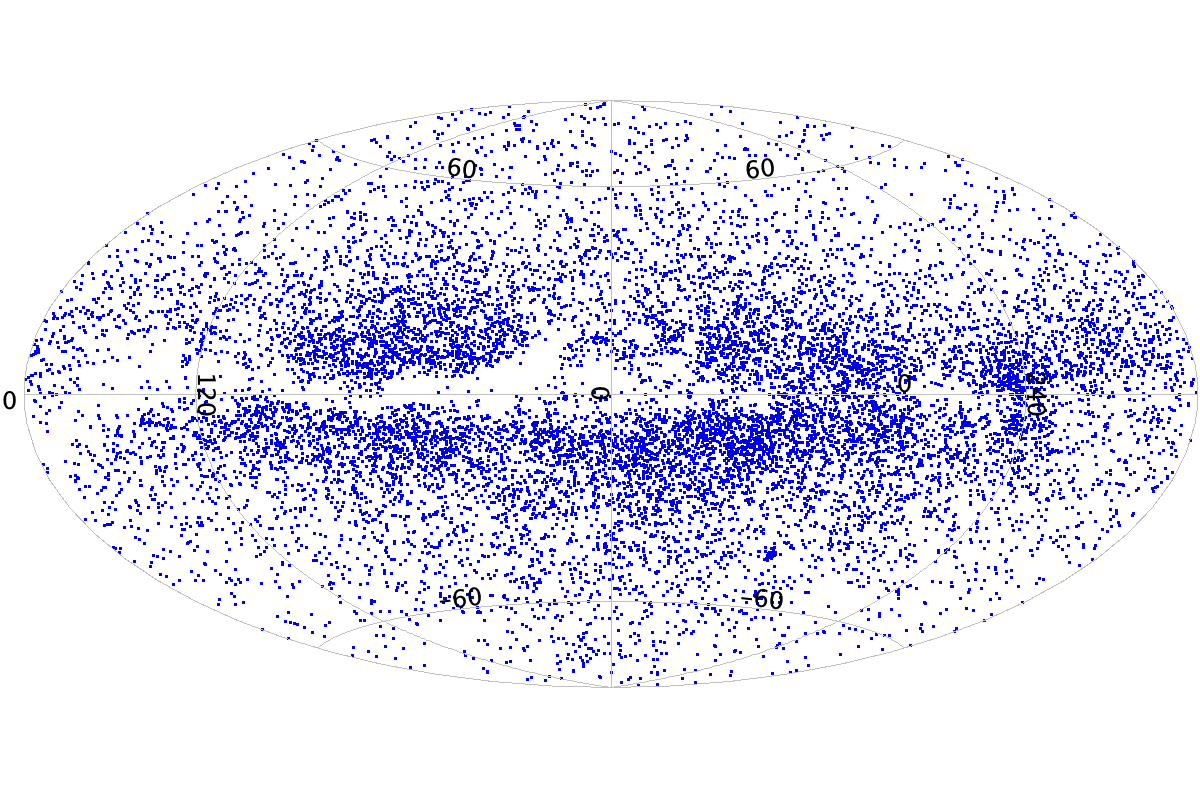}
  \includegraphics[width=\hsize]{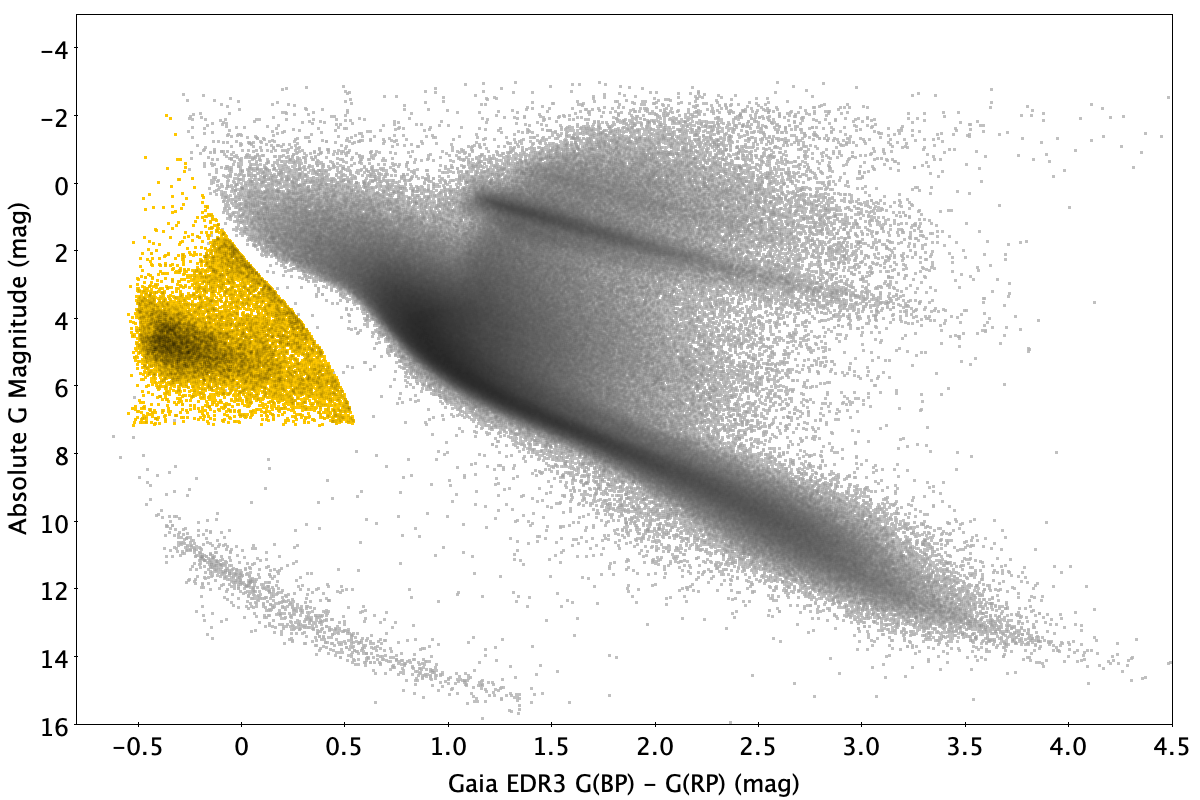}
 \caption{General properties of the Final Parallax Selection. Upper panel: Sky distribution of {\em Gaia} EDR3 Final Parallax Selection objects. Lower panel: Final Parallax Selection (yellow circles) shown on the full range {\em Gaia} CMD (grey dots).}
  \label{final_parallax_sky_distn}
  \end{figure*}

\begin{figure}
  \centering
  \includegraphics[width=\hsize]{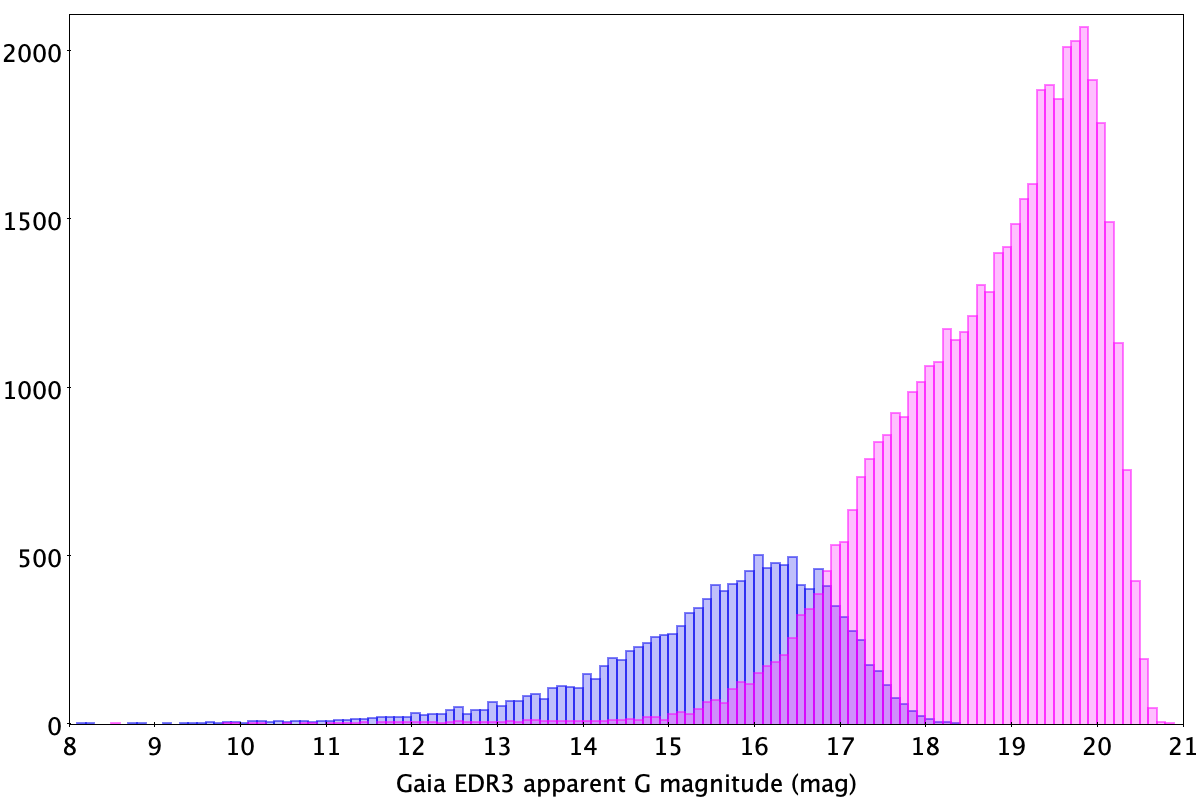}
 \caption{Distribution of {\em Gaia} EDR3 apparent g magnitude for the Final Parallax Selection (blue) and Final Proper Motion Selection (magenta).}
  \label{final_redPM_parallax_apmag_distn}
  \end{figure}

The {\em Gaia} EDR3 catalogue of hot subluminous star candidates consists of two parts (for a description of the catalogue columns, see Table~\ref{table:B1}). The Final Parallax Selection is a catalogue of 13,123 candidate objects found at distances of up to a few kiloparsecs with an apparent magnitude range from 7.0 up to 18.3 mag with the peak of the distribution at 16.2 mag (see Figure~\ref{final_redPM_parallax_apmag_distn}). We consider the Final Parallax Selection to be a full-sky catalogue (see Figure~\ref{final_parallax_sky_distn}) as there are no filtering or selection criteria based on Galactic latitude or similar. On the other hand, the crowded region filtering does remove some objects in the most crowded regions of the Galactic plane.

\begin{figure*}
  \centering
  \includegraphics[width=\hsize]{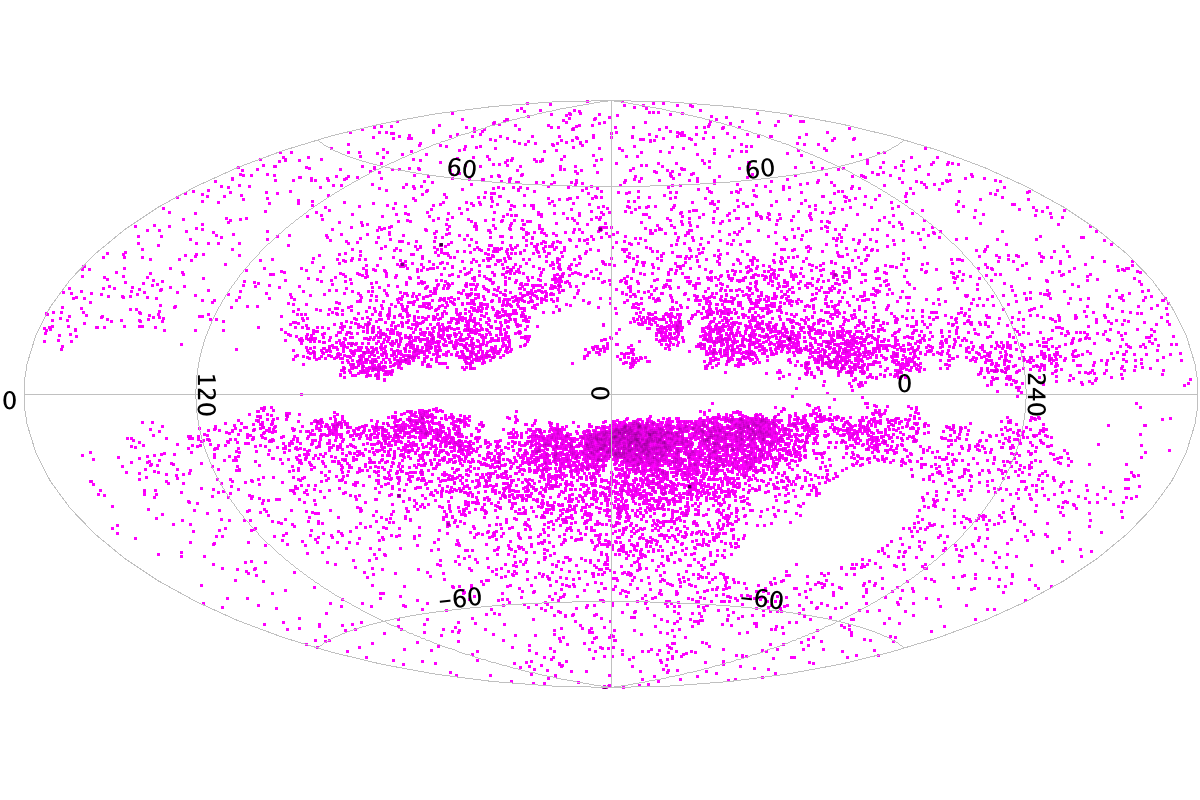}
 \caption{Sky distribution of {\em Gaia} EDR3 Final Proper Motion Selection objects.}
  \label{final_redpm_sky_distn}
  \end{figure*}

The Final Proper Motion Selection catalogue contains 41,822 candidate objects found at approximate distances of up to $\sim7\,{\rm kpc}$ with an apparent magnitude range of 8.6 mag up to 19.4 mag with the peak of the distribution at 18\,mag (see Figure~\ref{final_redPM_parallax_apmag_distn}). The crowded region filtering is stricter than that applied in the Final Parallax Selection and there is also a filter to cut out candidate objects in the direction of the Magellanic Clouds (see Figure~\ref{final_redpm_sky_distn}).

\begin{table*}[h!]
\centering
\begin{tabular}{ l c }
\hline\hline
Selection Criteria & Object Count \\ [0.5ex]
\hline
Parallax Selection Criteria: &\\
\hline
1. Parallax > 0, parallax error < 20$\%$, initial hot subdwarf CMD ranges & 3,213,406 \\
2. Position 1 + astrometric quality criteria & 3,213,398 \\
3. Position 2 + photometric quality criterion & 3,195,369 \\
4. Position 3 + CMD main sequence rejection criterion & 16,959 \\
5a. Position 4 + no apparent neighbours within 5 arcsec & 10,672 \\
5b. Position 4 + faint apparent neighbours within 5 arcsec & 1,147 \\
5c. Position 4 + bright apparent neighbours within 5 arcsec + strict hot subdwarf cutoff & 1,304 \\
6. Final parallax selection 5a + 5b + 5c & 13,123 \\
\hline
Proper Motion Selection Criteria& \\
\hline
7. Hot subdwarf colour-H\_G range, parallax error $\geq$20\% & 7,390,541 \\
8. Position 7 + proper motion error < 20\% & 6,860,074 \\
9. Position 8 + removal of the Magellanic Clouds direction & 4,856,823 \\
10. Position 9 + astrometric and photometric quality criteria & 4,784,233 \\
11. Position 10 + colour-H\_G main sequence rejection criterion & 92,409 \\
12a. Position 11 + no apparent neighbours within 5 arcsec & 66,393 \\
12b. Position 11 + faint apparent neighbours within 5 arcsec & 7,044 \\
13. Positions 12a+12b + white dwarf rejection criterion & 48,462 \\
\hline\hline
\end{tabular}
\caption{Table showing the effect of the hot subluminous star selection criteria for the parallax selection and the proper motion selection as outlined in Section 3. The definition of `bright' and `faint' apparent neighbours is described in Section 3.}
\label{table:3}
\end{table*}

%--------------------------------------------------------------------

\section{Summary and Conclusions}

In this work, we compiled two catalogues: The Known Hot Subdwarf Catalogue DR3 comprising 6,616 known hot subdwarfs with spectroscopic and photometric classifications; and the {\em Gaia} EDR3 Hot Subluminous Star catalogue comprising 13,123 hot subluminous star candidates with parallax error $<$~20$\%$ selected using absolute G magnitude and 48,462 hot subluminous star candidates with parallax error $\geq$~20$\%$ selected using reduced proper motion.

The known hot subdwarf catalogue DR3 was created using the methods used by \citet{geier20} but using the additional, recent discoveries as outlined in Section 2.1. The {\em Gaia} EDR3 Hot Subluminous Star catalogue was generated based on the procedures used in \citet{geier19} but modified to take into account the improvements in {\em Gaia} EDR3 photometry and astrometry compared to {\em Gaia} DR2. The data quality criteria were also modified to reflect the most recent published methods. The latest colour algorithms were also applied. The improved data quality in {\em Gaia} EDR3 also meant that we could improve the hot subluminous star candidate detection in the Galactic Plane and, in the Final Parallax Selection, in the direction of the Magellanic Clouds.

For the Final Parallax Selection, we were successful in using astrometric and photometric error excess measurements to discern between main sequence and hot subdwarf binary systems, unresolved hot subdwarf binary systems, and hot subdwarfs not in binary systems.

Our catalogues are the next step in the iterative process of creating a more complete, full-sky catalogue of hot subdwarf stars. It is known that the colour--magnitude diagram on its own is not sufficient to disentangle the diverse subtypes of hot subluminous stars and to distinguish them without any doubt from main sequence, BHB, and white dwarf stars. Spectroscopic follow-up is needed to learn more about this class of objects.

Our catalogue provides the selection criteria upon which future surveys can be based. In particular, upcoming large spectroscopic surveys (e.g. 4MOST, WEAVE, DESI) can construct their target lists using our catalogue. The target selection for the upcoming 4MOST hot subdwarf (HSD) consortium survey \citep{dejong19} to observe hot subluminous stars in the southern hemisphere will be based on our catalogues.

\begin{acknowledgements}

We thank Uli Heber and Harry Dawson for helpful commments and suggestions.

IP acknowledges funding by the UK’s Science and Technology Facilities Council (STFC), grant ST/T000406/1.

This research made use of TOPCAT, an interactive graphical viewer and editor for tabular data Taylor (\cite{taylor05}). This research made use of the SIMBAD database, operated at CDS, Strasbourg, France; the VizieR catalogue access tool, CDS, Strasbourg, France. Some of the data presented in this paper were obtained from the Mikulski Archive for Space Telescopes (MAST). STScI is operated by the Association of Universities for Research in Astronomy, Inc., under NASA contract NAS5-26555. Support for MAST for non-HST data is provided by the NASA Office of Space Science via grant NNX13AC07G and by other grants and contracts. This research has made use of the services of the ESO Science Archive Facility.

This work has made use of data from the European Space Agency (ESA) mission {\it Gaia} (https://www.cosmos.esa.int/gaia), processed by the {\it Gaia} Data Processing and Analysis Consortium (DPAC, https://www.cosmos.esa.int/web/gaia/dpac/consortium). Funding for the DPAC has been provided by national institutions, in particular the institutions participating in the {\it Gaia} Multilateral Agreement.

This publication makes use of data products from the Two Micron All Sky Survey, which is a joint project of the University of Massachusetts and the Infrared Processing and Analysis Center/California Institute of Technology, funded by the National Aeronautics and Space Administration and the National Science Foundation. Based on observations made with the NASA Galaxy Evolution Explorer. GALEX is operated for NASA by the California Institute of Technology under NASA contract NAS5-98034. This research has made use of the APASS database, located at the AAVSO web site. Funding for APASS has been provided by the Robert Martin Ayers Sciences Fund. The Guide Star catalogue-II is a joint project of the Space Telescope Science Institute and the Osservatorio Astronomico di Torino. Space Telescope Science Institute is operated by the Association of Universities for Research in Astronomy, for the National Aeronautics and Space Administration under contract NAS5-26555. The participation of the Osservatorio Astronomico di Torino is supported by the Italian Council for Research in Astronomy. Additional support is provided by European Southern Observatory, Space Telescope European Coordinating Facility, the International GEMINI project and the European Space Agency Astrophysics Division.

Based on observations obtained as part of the VISTA Hemisphere Survey, ESO Program, 179.A-2010 (PI: McMahon). This publication has made use of data from the VIKING survey from VISTA at the ESO Paranal Observatory, programme ID 179.A-2004. Data processing has been contributed by the VISTA Data Flow System at CASU, Cambridge and WFAU, Edinburgh. Based on data products from observations made with ESO Telescopes at the La Silla Paranal Observatory under program ID 177.A 3011(A,B,C,D,E.F). Based on data products from observations made with ESO Telescopes at the La Silla Paranal Observatory under programme IDs 177.A-3016, 177.A-3017 and 177.A-3018, and on data products produced by Target/OmegaCEN, INAF-OACN, INAF-OAPD and the KiDS production team, on behalf of the KiDS consortium. OmegaCEN and the KiDS production team acknowledge support by NOVA and NWO-M grants. Members of INAF-OAPD and INAF-OACN also acknowledge the support from the Department of Physics \& Astronomy of the University of Padova, and of the Department of Physics of Univ. Federico II (Naples). This publication makes use of data products from the Wide-field Infrared Survey Explorer, which is a joint project of the University of California, Los Angeles, and the Jet Propulsion Laboratory/California Institute of Technology, and NEOWISE, which is a project of the Jet Propulsion Laboratory/California Institute of Technology. WISE and NEOWISE are funded by the National Aeronautics and Space Administration.

The Pan-STARRS1 Surveys (PS1) and the PS1 public science archive have been made possible through contributions by the Institute for Astronomy, the University of Hawaii, the Pan-STARRS Project Office, the Max-Planck Society and its participating institutes, the Max Planck Institute for Astronomy, Heidelberg and the Max Planck Institute for Extraterrestrial Physics, Garching, The Johns Hopkins University, Durham University, the University of Edinburgh, the Queen's University Belfast, the Harvard-Smithsonian Center for Astrophysics, the Las Cumbres Observatory Global Telescope Network Incorporated, the National Central University of Taiwan, the Space Telescope Science Institute, the National Aeronautics and Space Administration under Grant No. NNX08AR22G issued through the Planetary Science Division of the NASA Science Mission Directorate, the National Science Foundation Grant No. AST-1238877, the University of Maryland, Eotvos Lorand University (ELTE), the Los Alamos National Laboratory, and the Gordon and Betty Moore Foundation.

The national facility capability for SkyMapper has been funded through ARC LIEF grant LE130100104 from the Australian Research Council, awarded to the University of Sydney, the Australian National University, Swinburne University of Technology, the University of Queensland, the University of Western Australia, the University of Melbourne, Curtin University of Technology, Monash University and the Australian Astronomical Observatory. SkyMapper is owned and operated by The Australian National University's Research School of Astronomy and Astrophysics. The survey data were processed and provided by the SkyMapper Team at ANU. The SkyMapper node of the All-Sky Virtual Observatory (ASVO) is hosted at the National Computational Infrastructure (NCI). Development and support the SkyMapper node of the ASVO has been funded in part by Astronomy Australia Limited (AAL) and the Australian Government through the Commonwealth's Education Investment Fund (EIF) and National Collaborative Research Infrastructure Strategy (NCRIS), particularly the National eResearch Collaboration Tools and Resources (NeCTAR) and the Australian National Data Service Projects (ANDS).

Guoshoujing Telescope (the Large Sky Area Multi-Object Fiber Spectroscopic Telescope LAMOST) is a National Major Scientific Project built by the Chinese Academy of Sciences. Funding for the project has been provided by the National Development and Reform Commission. LAMOST is operated and managed by the National Astronomical Observatories, Chinese Academy of Sciences.

Funding for the SDSS and SDSS-II has been provided by the Alfred P. Sloan Foundation, the Participating Institutions, the National Science Foundation, the U.S. Department of Energy, the National Aeronautics and Space Administration, the Japanese Monbukagakusho, the Max Planck Society, and the Higher Education Funding Council for England. The SDSS Web Site is http://www.sdss.org/. The SDSS is managed by the Astrophysical Research Consortium for the Participating Institutions. The Participating Institutions are the American Museum of Natural History, Astrophysical Institute Potsdam, University of Basel, University of Cambridge, Case Western Reserve University, University of Chicago, Drexel University, Fermilab, the Institute for Advanced Study, the Japan Participation Group, Johns Hopkins University, the Joint Institute for Nuclear Astrophysics, the Kavli Institute for Particle Astrophysics and Cosmology, the Korean Scientist Group, the Chinese Academy of Sciences (LAMOST), Los Alamos National Laboratory, the Max-Planck-Institute for Astronomy (MPIA), the Max-Planck-Institute for Astrophysics (MPA), New Mexico State University, Ohio State University, University of Pittsburgh, University of Portsmouth, Princeton University, the United States Naval Observatory, and the University of Washington. 

Funding for SDSS-III has been provided by the Alfred P. Sloan Foundation, the Participating Institutions, the National Science Foundation, and the U.S. Department of Energy Office of Science. The SDSS-III web site is http://www.sdss3.org/. SDSS-III is managed by the Astrophysical Research Consortium for the Participating Institutions of the SDSS-III Collaboration including the University of Arizona, the Brazilian Participation Group, Brookhaven National Laboratory, University of Cambridge, Carnegie Mellon University, University of Florida, the French Participation Group, the German Participation Group, Harvard University, the Instituto de Astrofisica de Canarias, the Michigan State/Notre Dame/JINA Participation Group, Johns Hopkins University, Lawrence Berkeley National Laboratory, Max Planck Institute for Astrophysics, Max Planck Institute for Extraterrestrial Physics, New Mexico State University, New York University, Ohio State University, Pennsylvania State University, University of Portsmouth, Princeton University, the Spanish Participation Group, University of Tokyo, University of Utah, Vanderbilt University, University of Virginia, University of Washington, and Yale University. 

Funding for the Sloan Digital Sky 
Survey IV has been provided by the 
Alfred P. Sloan Foundation, the U.S. 
Department of Energy Office of 
Science, and the Participating 
Institutions. 

SDSS-IV acknowledges support and 
resources from the Center for High 
Performance Computing at the 
University of Utah. The SDSS 
website is www.sdss.org.

SDSS-IV is managed by the 
Astrophysical Research Consortium 
for the Participating Institutions 
of the SDSS Collaboration including 
the Brazilian Participation Group, 
the Carnegie Institution for Science, 
Carnegie Mellon University, Center for 
Astrophysics | Harvard \& 
Smithsonian, the Chilean Participation 
Group, the French Participation Group, 
Instituto de Astrof\'isica de 
Canarias, The Johns Hopkins 
University, Kavli Institute for the 
Physics and Mathematics of the 
Universe (IPMU) / University of 
Tokyo, the Korean Participation Group, 
Lawrence Berkeley National Laboratory, 
Leibniz Institut f\"ur Astrophysik 
Potsdam (AIP), Max-Planck-Institut 
f\"ur Astronomie (MPIA Heidelberg), 
Max-Planck-Institut f\"ur 
Astrophysik (MPA Garching), 
Max-Planck-Institut f\"ur 
Extraterrestrische Physik (MPE), 
National Astronomical Observatories of 
China, New Mexico State University, 
New York University, University of 
Notre Dame, Observat\'ario 
Nacional / MCTI, The Ohio State 
University, Pennsylvania State 
University, Shanghai 
Astronomical Observatory, United 
Kingdom Participation Group, 
Universidad Nacional Aut\'onoma 
de M\'exico, University of Arizona, 
University of Colorado Boulder, 
University of Oxford, University of 
Portsmouth, University of Utah, 
University of Virginia, University 
of Washington, University of 
Wisconsin, Vanderbilt University, 
and Yale University.

\end{acknowledgements}

\clearpage
\begin{onecolumn}
\begin{appendix}

\section{Catalogue of known hot subdwarfs DR3}

\begin{longtable}{llll}
\caption{\label{table:A1} Catalogue columns}\\
\hline\hline
\noalign{\smallskip}
Column & Format & Description & Unit \\
\noalign{\smallskip}
\hline
\noalign{\smallskip}
NAME & A30 & Target name & \\
GAIA\_DESIG & A30 & Gaia designation & \\
RA & F10.6 & Right ascension (J2000) & deg \\
DEC & F10.6 & Declination (J2000) & deg \\
GLON & F10.6 & Galactic longitude & deg \\
GLAT & F10.6 & Galactic latitude & deg \\
SPEC\_CLASS & A15 & Spectroscopic classification & \\
SPEC\_SIMBAD & A15 & Spectroscopic classification from SIMBAD & \\
COLOUR\_SDSS & A10 & Colour classification SDSS & \\
COLOUR\_APASS & A10 & Colour classification GALEX/APASS & \\
COLOUR\_PS1 & A10 & Colour classification GALEX/PS1 & \\
COLOUR\_SKYM & A10 & Colour classification SkyMapper & \\
PLX & F8.4 & Gaia parallax & mas \\
PLX\_ZP & F8.4 & Zero-point-corrected Gaia parallax & mas \\
e\_PLX & F8.4 & Error on PLX & mas \\
M\_G   & F8.4 & Absolute magnitude in G-band & mag \\
G\_GAIA & F6.3 & Gaia G-band magnitude & mag \\
e\_G\_GAIA & F6.3 & Error on G\_GAIA & mag \\
BP\_GAIA & F6.3 & Gaia BP-band magnitude & mag \\
e\_BP\_GAIA & F6.3 & Error on BP\_GAIA & mag \\
RP\_GAIA & F6.3 & Gaia RP-band magnitude & mag \\
e\_RP\_GAIA & F6.3 & Error on RP\_GAIA & mag \\
PMRA\_GAIA & F7.3 & Gaia proper motion $\mu_{\rm \alpha}\cos{\rm \delta}$ & ${\rm mas\,yr^{-1}}$ \\
e\_PMRA\_GAIA & F7.3 & Error on PMRA\_GAIA & ${\rm mas\,yr^{-1}}$ \\
PMDEC\_GAIA & F7.3 & Gaia proper motion $\mu_{\rm \delta}$ & ${\rm mas\,yr^{-1}}$ \\
e\_PMDEC\_GAIA & F7.3 & Error on PMDEC\_GAIA & ${\rm mas\,yr^{-1}}$ \\
RV\_SDSS & F5.1 & Radial velocity SDSS & ${\rm km\,s^{-1}}$ \\
e\_RV\_SDSS & F5.1 & Error on RV\_SDSS & ${\rm km\,s^{-1}}$ \\
RV\_LAMOST & F5.1 & Radial velocity LAMOST & ${\rm km\,s^{-1}}$ \\
e\_RV\_LAMOST & F5.1 & Error on RV\_LAMOST & ${\rm km\,s^{-1}}$ \\
TEFF & F8.1 & Effective temperature & K \\
e\_TEFF & F8.1 & Error on T\_EFF & K \\
LOG\_G & F4.2 & Log surface gravity (gravity in ${\rm cm\,s^{-2}}$) & dex \\
e\_LOG\_G & F.4.2 & Error on LOG\_G & dex \\
LOG\_Y & F5.2 & Log helium abundance $n({\rm He})/n({\rm H})$ & dex \\
e\_LOG\_Y & F5.2 & Error on LOG\_Y & dex \\
PARAMS\_REF & A20 & Reference for atmospheric parameters (Bibcode)\footnote{References from the literature: \citet{tomley70,greenstein73,giddings81,hunger81,heber84,heber86a,heber86b,heber86c,heber86d,heber87,heber02,heber03,husfeld89,dreizler90,moehler90,bixler91,rauch91,rauch14,viton91,jeffery92,jeffery13,jeffery21,rauch93,theissen93,saffer94,saffer97,thejll94,bauer95,rauch95,lanz97,lemke97,dreizler98,koen98,napiwotzki99,edelmann99,wood99,burleigh00,maxted01,ramspeck01,maxted02,ahmad03,edelmann03,morales03,werner04a,werner04b,werner05,werner14,werner15,werner22a,werner22b,lisker05,otoole06,huegelmeyer06,przybilla06,monibidin07,monibidin09,monibidin12,stroeer07,charpinet08,charpinet10,geier08,geier10,geier11,geier13,geier13a,geier13b,geier14,geier15b,geier17b,geier22a,geier22b,fontaine08,hirsch09,todt09,for10,vangrootel10,barlow10,barlow13,jeffery10,naslim10,oestensen10a,oestensen10b,oestensen10c,oestensen14,klepp11,latour11,latour15,latour18a,tillich11,bloemen11,copperwheat11,koen11,herald11,miszalski12,nemeth12,nemeth16,vos12,vos13,almeida12,baran12,baran16,baran19,verbeek12,ziegler12,schaffenroth13,schaffenroth19,schaffenroth21,frew14,reindl14,kupfer15,kupfer17a,kupfer17b,kupfer20,kupfer22,kepler16,kepler19,reindl16,reindl17,reindl20,derekas15,chayer15,demarco15,aller15,bachulski16,luo16,luo19,luo21,ziegerer17,hillwig17,holdsworth17,lei18,lei19,lei20,gvaramadze19,ratzloff19,ratzloff20,kilkenny19,loebling19,bell19,loebling20,hogg20,silvotti21,silvotti22,pelisoli21,dorsch22}} &  \\
EB-V & F6.4 & Interstellar reddening E(B-V) & mag \\
e\_EB-V & F6.4 & Error on EB-V & mag \\
AV & F6.4 & Interstellar extinction A$_{\rm V}$ & mag \\
FUV\_GALEX & F6.3 & GALEX FUV-band magnitude & mag \\
e\_FUV\_GALEX & F6.3 & Error on FUV\_GALEX & mag \\
NUV\_GALEX & F6.3 & GALEX NUV-band magnitude & mag \\
e\_NUV\_GALEX & F6.3 & Error on NUV\_GALEX & mag \\
V\_APASS & F6.3 & APASS V-band magnitude & mag \\
e\_V\_APASS & F6.3 & Error on V\_APASS & mag \\
B\_APASS & F6.3 & APASS B-band magnitude & mag \\
e\_B\_APASS & F6.3 & Error on V\_APASS & mag \\
g\_APASS & F6.3 & APASS g-band magnitude & mag \\
e\_g\_APASS & F6.3 & Error on g\_APASS & mag \\
r\_APASS & F6.3 & APASS r-band magnitude & mag \\
e\_r\_APASS & F6.3 & Error on r\_APASS & mag \\
i\_APASS & F6.3 & APASS i-band magnitude & mag \\
e\_i\_APASS & F6.3 & Error on i\_APASS & mag \\
u\_SDSS & F6.3 & SDSS u-band magnitude & mag \\
e\_u\_SDSS & F6.3 & Error on u\_SDSS & mag \\
g\_SDSS & F6.3 & SDSS g-band magnitude & mag \\
e\_g\_SDSS & F6.3 & Error on g\_SDSS & mag \\
r\_SDSS & F6.3 & SDSS r-band magnitude & mag \\
e\_r\_SDSS & F6.3 & Error on r\_SDSS & mag \\
i\_SDSS & F6.3 & SDSS i-band magnitude & mag \\
e\_i\_SDSS & F6.3 & Error on i\_SDSS & mag \\
z\_SDSS & F6.3 & SDSS z-band magnitude & mag \\
e\_z\_SDSS & F6.3 & Error on z\_SDSS & mag \\
u\_VST & F6.3 & VST surveys (ATLAS, KiDS) u-band magnitude & mag \\
e\_u\_VST & F6.3 & Error on u\_VST & mag \\
g\_VST & F6.3 & VST surveys (ATLAS, KiDS) g-band magnitude & mag \\
e\_g\_VST & F6.3 & Error on g\_VST & mag \\
r\_VST & F6.3 & VST surveys (ATLAS, KiDS) r-band magnitude & mag \\
e\_r\_VST & F6.3 & Error on r\_VST & mag \\
i\_VST & F6.3 & VST surveys (ATLAS, KiDS) i-band magnitude & mag \\
e\_i\_VST & F6.3 & Error on i\_VST & mag \\
z\_VST & F6.3 & VST surveys (ATLAS, KiDS) z-band magnitude & mag \\
e\_z\_VST & F6.3 & Error on z\_VST & mag \\
u\_SKYM & F6.3 & SkyMapper u-band magnitude & mag \\
e\_u\_SKYM & F6.3 & Error on u\_SKYM & mag \\
v\_SKYM & F6.3 & SkyMapper v-band magnitude & mag \\
e\_v\_SKYM & F6.3 & Error on v\_SKYM & mag \\
g\_SKYM & F6.3 & SkyMapper g-band magnitude & mag \\
e\_g\_SKYM & F6.3 & Error on g\_SKYM & mag \\
r\_SKYM & F6.3 & SkyMapper r-band magnitude & mag \\
e\_r\_SKYM & F6.3 & Error on r\_SKYM & mag \\
i\_SKYM & F6.3 & SkyMapper i-band magnitude & mag \\
e\_i\_SKYM & F6.3 & Error on i\_SKYM & mag \\
z\_SKYM & F6.3 & SkyMapper z-band magnitude & mag \\
e\_z\_SKYM & F6.3 & Error on z\_SKYM & mag \\
g\_PS1 & F7.4 & PS1 g-band magnitude & mag \\
e\_g\_PS1 & F7.4 & Error on g\_PS1 & mag \\
r\_PS1 & F7.4 & PS1 r-band magnitude & mag \\
e\_r\_PS1 & F7.4 & Error on r\_PS1 & mag \\
i\_PS1 & F7.4 & PS1 i-band magnitude & mag \\
e\_i\_PS1 & F7.4 & Error on i\_PS1 & mag \\
z\_PS1 & F7.4 & PS1 z-band magnitude & mag \\
e\_z\_PS1 & F7.4 & Error on z\_PS1 & mag \\
y\_PS1 & F7.4 & PS1 y-band magnitude & mag \\
e\_y\_PS1 & F7.4 & Error on y\_PS1 & mag \\
J\_2MASS & F6.3 & 2MASS J-band magnitude & mag \\
e\_J\_2MASS & F6.3 & Error on J\_2MASS & mag \\
H\_2MASS & F6.3 & 2MASS H-band magnitude & mag \\
e\_H\_2MASS & F6.3 & Error on H\_2MASS & mag \\
K\_2MASS & F6.3 & 2MASS K-band magnitude & mag \\
e\_K\_2MASS & F6.3 & Error on K\_2MASS & mag \\
Y\_UKIDSS & F6.3 & UKIDSS Y-band magnitude & mag \\
e\_Y\_UKIDSS & F6.3 & Error on Y\_UKIDSS & mag \\
J\_UKIDSS & F6.3 & UKIDSS J-band magnitude & mag \\
e\_J\_UKIDSS & F6.3 & Error on J\_UKIDSS & mag \\
H\_UKIDSS & F6.3 & UKIDSS H-band magnitude & mag \\
e\_H\_UKIDSS & F6.3 & Error on H\_UKIDSS & mag \\
K\_UKIDSS & F6.3 & UKIDSS K-band magnitude & mag \\
e\_K\_UKIDSS & F6.3 & Error on K\_UKIDSS & mag \\
Z\_VISTA & F6.3 & VISTA surveys (VHS, VIKING) Z-band magnitude & mag \\
e\_Z\_VISTA & F6.3 & Error on Z\_VISTA & mag \\
Y\_VISTA & F6.3 & VISTA surveys (VHS, VIKING) Y-band magnitude & mag \\
e\_Y\_VISTA & F6.3 & Error on Y\_VISTA & mag \\
J\_VISTA & F6.3 & VISTA surveys (VHS, VIKING) J-band magnitude & mag \\
e\_J\_VISTA & F6.3 & Error on J\_VISTA & mag \\
H\_VISTA & F6.3 & VISTA surveys (VHS, VIKING) H-band magnitude & mag \\
e\_H\_VISTA & F6.3 & Error on H\_VISTA & mag \\
Ks\_VISTA & F6.3 & VISTA surveys (VHS, VIKING) Ks-band magnitude & mag \\
e\_Ks\_VISTA & F6.3 & Error on Ks\_VISTA & mag \\
W1 & F6.3 & WISE W1-band magnitude & mag \\
e\_W1 & F6.3 & Error on W1 & mag \\
W2 & F6.3 & WISE W2-band magnitude & mag \\
e\_W2 & F6.3 & Error on W2 & mag \\
W3 & F6.3 & WISE W3-band magnitude & mag \\
e\_W3 & F6.3 & Error on W3 & mag \\
W4 & F6.3 & WISE W4-band magnitude & mag \\
e\_W4 & F6.3 & Error on W4 & mag \\
\noalign{\smallskip}
\hline\hline
\end{longtable}

\section{Gaia EDR3 catalogues of hot subluminous star candidates}

\begin{longtable}{llll}
\caption{\label{table:B1} Catalogue columns}\\
\hline\hline
\noalign{\smallskip}
Column & Format & Description & Unit \\
\noalign{\smallskip}
\hline
\noalign{\smallskip}
source\_id & I19 & Gaia EDR3 source identifier & - \\
ra & F10.6 & Gaia EDR3 Right ascension & deg \\
dec & F10.6 & Gaia EDR3 Declination & deg \\
l & F10.6 & Galactic longitude & deg \\
b & F10.6 & Galactic latitude & deg \\
parallax & F8.4 & Gaia parallax $\bar{\omega}$ & mas \\
parallax\_error & F8.4 & Error on parallax & mas \\
abs\_g\_mag & F8.4 & Absolute magnitude in G-band & mag \\
phot\_g\_mean\_mag & F6.3 & Gaia apparent G magnitude & mag \\
bp\_rp & F6.3 & Gaia $G_{BP}$ - $G_{RP}$ magnitude & mag \\
phot\_bp\_rp\_excess\_factor\_corrected & F6.3 & Corrected Gaia $G_{BP}$ - $G_{RP}$ excess factor & - \\
pmra & F7.3 & Gaia proper motion $\mu_{\rm \alpha}\cos{\rm \delta}$ & ${\rm mas\,yr^{-1}}$ \\
pmra\_error & F7.3 & Error on pmra & ${\rm mas\,yr^{-1}}$ \\
pmdec & F7.3 & Gaia proper motion $\mu_{\rm \delta}$ & ${\rm mas\,yr^{-1}}$ \\
pmdec\_error & F7.3 & Error on pmdec & ${\rm mas\,yr^{-1}}$ \\
pm & F7.3 & Gaia proper motion $\mu$ & ${\rm mas\,yr^{-1}}$ \\
pm\_error & F7.3 & Error on proper motion & ${\rm mas\,yr^{-1}}$ \\
ruwe & F7.5 & Gaia EDR3 renormalised unit weight error & - \\
reduced\_proper\_motion & F7.3 & Reduced proper motion H & mag \\
excess\_flux\_error & F7.3 & Excess flux error & - \\
parallax\_selection\_flag & I1 & Parallax selection catalogue candidate & - \\
proper\_motion\_selection\_flag & I1 & Proper motion selection catalogue candidate & - \\

\noalign{\smallskip}
\hline\hline
\end{longtable}

\end{appendix}

\end{onecolumn}

\end{document}